\documentclass[useAMS,usenatbib]{mn2e}
\usepackage{rotating}
\usepackage{subfigure}
\RequirePackage{setspace}
\usepackage{graphicx}
\usepackage{fancyhdr}
\usepackage{natbib}
\usepackage{amsmath}
\usepackage{amssymb}
\usepackage{aas_macros}
\usepackage{epstopdf}
\usepackage{float}
\usepackage[normalem]{ulem}

\title[FIR emission from high-$z$ galaxies]{The detection of FIR emission from high redshift star-forming galaxies in the ECDF-S}
\author[L. J. M. Davies et. al.]{L. J. M. Davies$^{1}$\thanks{E-mail:
Luke.Davies@bristol.ac.uk}, M. N. Bremer$^{1}$, E. R. Stanway$^{2}$, M. D. Lehnert$^{3}$\\
$^{1}$Department of Physics, University of Bristol, H.H. Wills Physics Laboratory, Tyndall Avenue, Bristol, BS8 1TL, UK\\
$^{2}$Department of Physics, University of Warwick, Gibbet Hill Road, Coventry, CV4 7AL, UK\\
$^{3}$Institut dÕAstrophysique de Paris, UniversitŽ Pierre et Marie Curie/CNRS,98 bis Bd Arago, 75014 Paris, France}

\begin{document}

\date{Draft: Feb 2013}

\pagerange{\pageref{firstpage}--\pageref{lastpage}} \pubyear{2010}

\maketitle

\begin{abstract}

We have used the Large Apex Bolometer Camera (LABOCA) Survey of the Extended Chandra Deep Field South (LESS) to investigate rest-frame FIR emission from typical star-forming systems (Lyman Break Galaxies, LBGs, and Lyman-$\alpha$ emitters, LAEs) at redshift 3, 4, 4.5 and 5 (922, 68, 46, and 20 sources respectively). We initially concentrate on LBGs at $z\sim3$ and select three subsamples on stellar mass (rest-frame optical-brightest, M$_{*}>10.^{10.25}$M$_{\odot}$), extinction corrected star-formation (assuming $\beta_{UV}=-2$ and applying a dust attenuation correction, SFR$_{tot} > 6.7$M$_{\odot}$\,yr$^{-1}$) and rest-frame UV-magnitude (representing a typical Lyman-break selection with $R <$ 24.43). We produce composite 870$\mu$m images of the typical source in our subsamples, obtaining  $\sim4\sigma$ detections (0.61\,mJy and 0.35\,mJy and 0.37\,mJy respectively) and suggesting a correlation between FIR luminosity and stellar mass. We apply a similar procedure to our full samples at $z\sim3$, 4, 4.5 and 5 and do not obtain detections - a result that is consistent with a simple scaling between FIR luminosity and stellar mass. In order to constrain the FIR SED of these systems we explore their emission at multiple wavelengths spanning the peak of dust emission at $z\sim3$ using the $Herschel$ SPIRE observations of the field. We obtain detections at multiple wavelengths of both our stellar mass and UV-magnitude selected samples, and find a best-fit SED with dust temperatures in the $\sim33-41$K range. We calculate FIR luminosity, obscured SFRs and dust masses and find that a significant fraction of star-formation in these systems is obscured. Interestingly, our sample selected on extinction corrected SFR does not display the large FIR fluxes predicted from its red UV-spectral slope. This suggests that the method of assuming an intrinsic UV-slope and correcting for dust attenuation may be invalid for this sample - and that these are not in fact the most actively star-forming systems. All of our $z\sim3$ samples fall on the `main sequence' of star-forming galaxies at $z\sim3$ and our detected subsamples are likely to represent the high obscuration end of the LBGs population at their epoch. We compare the FIR properties of our subsamples with various other galaxy populations, finding that our stellar mass selected sample shows some similar FIR characteristics to SMGs at the same epoch and therefore potentially represents the low FIR luminosity end of the high redshift FIR luminosity function.

\end{abstract}


\section{Introduction}
\label{Into}

Lyman Break Galaxies \citep[LBGs, $e.g.$][]{Steidel95} form a substantial fraction of the
observed high redshift ($z\gtrsim3$) galaxy population
\citep[e.g][]{Steidel95, vanzella09,Douglas09,Douglas10}. Primarily identified via bright UV-emission emission which arises from hot, young O and B stars in relatively unobscured regions. The properties of these systems have been extensively studied via their rest-frame UV spectra and UV-optical spectral energy distributions \citep[e.g][]{Shapley01, Rigopoulou06, Verma07, Stark09}. Given that in the past two decades LBG samples have dominated observational studies of galaxies at $z>3$ it is unfortunate that comparatively little work has been carried out to explore the properties of their non-optical/UV bright components (interstellar gas and cool dust), which are necessary for a more complete picture of these early star-forming systems. If we wish to build a comprehensive understanding of star-formation activity and galactic evolution in high redshift systems we must observe their complete baryonic budget of stellar material, dust and molecular gas. Through a detailed comparison of the stellar, molecular gas and dust fractions we can infer their star formation history and potential fate - thereby investigating their importance to the evolution of galaxies in general. In our previous studies we have investigated the molecular gas and dust content of $z\sim5$ LBGs \cite{Stanway08, Stanway10, Davies10, Davies12}, here we expand this work to consider the dust content of LBGs over a range of epochs.

Several observational studies have attempted to constrain the dust content of $z\gtrsim3$ star-forming galaxies \cite[$e.g.$][]{Chapman00, Webb03, Carilli07, Chapman09, Stanway10, Davies12, Lee12, Oteo13}. In combination, these works have determined that typical LBGs have relatively faint far-infrared luminosities (L$_{\mathrm{FIR}}\lesssim$ few$\times10^{11}$L$_{\odot}$) and therefore low dust masses (M$_{\mathrm{dust}}\lesssim $ few$ \times10^{8}$M$_{\odot}$) - for reasonable assumptions of dust temperature and spectral energy distribution. While these studies have produced relatively tight constraints on the FIR emission from these sources, the small number of galaxies targeted, up to $\sim140$ photometrically selected objects at $z\sim3$ \citep{Webb03} and $\sim20$ spectroscopically confirmed sources at $z\sim5$ \citep{Stanway10,Davies12}, limits the depth of any combined image used to determine their typical FIR properties ($\gtrsim0.5$mJy/beam at 870\,$\mu$m). Some individual detections have been achieved by targeting highly lensed LBGs at $z\sim3$, typically obtaining unlensed 850\,$\mu$m fluxes of $<$0.8mJy \citep[$e.g.$][ and see Chapman \& Casey (2009) for a summary]{Baker01, Borys04, Chapman02, Coppin07, Conley11, Kneib05, Negrello10,Siana09}. In addition, \cite{Baker01} use SED fitting of a single, highly lensed LBG at $z=2.7$ (cB58) to obtain a dust temperature of T\,=\,33\,K. Although these detections place interesting constraints on the FIR emission from LBGs it is unclear as to whether or not these individual detections are representative of the whole LBG population.      

While the typical LBG remains undetected, greater success has been obtained through targeting atypically massive  and UV-bright LBGs. Recently, \cite{magdis101} claim a detection in a stacked sample of the most massive LBGs in the Great Observatories Origins Deep Survey - North (GOODS-N) field and obtain a 0.41\,mJy flux at 1.1mm. For any reasonable assumption of dust SED, this suggests that the relatively massive, rest frame optically selected, sources should be detectable at at 870$\mu$m at a $\sim0.6$\,mJy level. At $z\sim4$, \cite{Lee12} investigated FIR emission from the most actively star-forming systems in the Bootes field ($\sim2000$ sources). They split their sample into three magnitude bins and obtain 350 and 500$\mu$m detections in the two highest luminosity bins - finding that the most UV-luminous systems are the most FIR bright.   

However, without multiple direct FIR detections, spanning the peak of the dust emission for the typical source at these redshifts, we can say little about the properties of the obscured material in these systems. Prior to this work the best estimates of the dust SED shape of $z\sim3$ LBGs were either loosely constrained by limits or single wavelength detections \cite[$e.g.$][]{Magdis10}, extrapolated from sub-mm bright sources at similar epochs \cite[$e.g.$ sub-mm galaxies, see discussion in][]{Davies12} or modelled from low redshift analogous sources \cite[such as blue compact dwarfs,][]{Dale07}. Hence, a direct determination of the dust SED for high redshift sources is required in order to successfully constrain the dust properties of these systems and investigate their obscured stellar populations.   

In our recent study \citep{Davies12} we investigated the FIR emission from a small sample of spectroscopically confirmed $z\sim5$ LBGs. In this new work we use the the LABOCA Survey of the Extended Chandra Deep Field South (LESS) to expand this study and investigate the FIR properties of a large number of both spectroscopically confirmed and photometrically identified LBGs over a range of redshifts. Utilising the deep $870\mu$m observations over the relatively large area of the ECDF-S we produce a comprehensive study of FIR emission from high redshift unobscured star-forming galaxies. We select subsamples of the most massive, those predicted to be the most actively star-forming (therefore, potentially the most FIR bright) and most rest-frame UV-bright $z\sim3$ galaxies in the field and obtain detections of the typical source in each sample. We identify two of our samples in the $Herschel$ maps of the field at 250, 350 and 500\,$\mu$m allowing us to constrain the $z\sim3$ LBG dust SED. Therefore, we make the first reliable estimates for the  FIR luminosity, obscured star-formation, dust mass, and dust temperature in these early star-forming systems.     

The technical analysis which is undertaken in the paper is comparable to that outlined in \cite{Davies12}, albeit on a wider range of samples. Hence, we shall only briefly discuss our methods and refer the reader to \cite{Davies12} for any further details. We note that, unless otherwise stated, when discussing UV and FIR emission we refer to emission in the rest-frame of the galaxy. Throughout this paper all optical magnitudes are quoted in the AB
system \citep{Oke83}, and the cosmology used is
\textit{H}$_{0}$\,=\,70kms$^{-1}$ Mpc$^{-1}$,
$\Omega_{\Lambda}$\,=\,0.7 and $\Omega_{M}$\,=\,0.3.

\section{Data Sets and Source Selection}
\label{sec:data}

\begin{figure*}
\begin{center}

\subfigure[   $z\sim$2.8-3.6]{\includegraphics[scale=0.42]{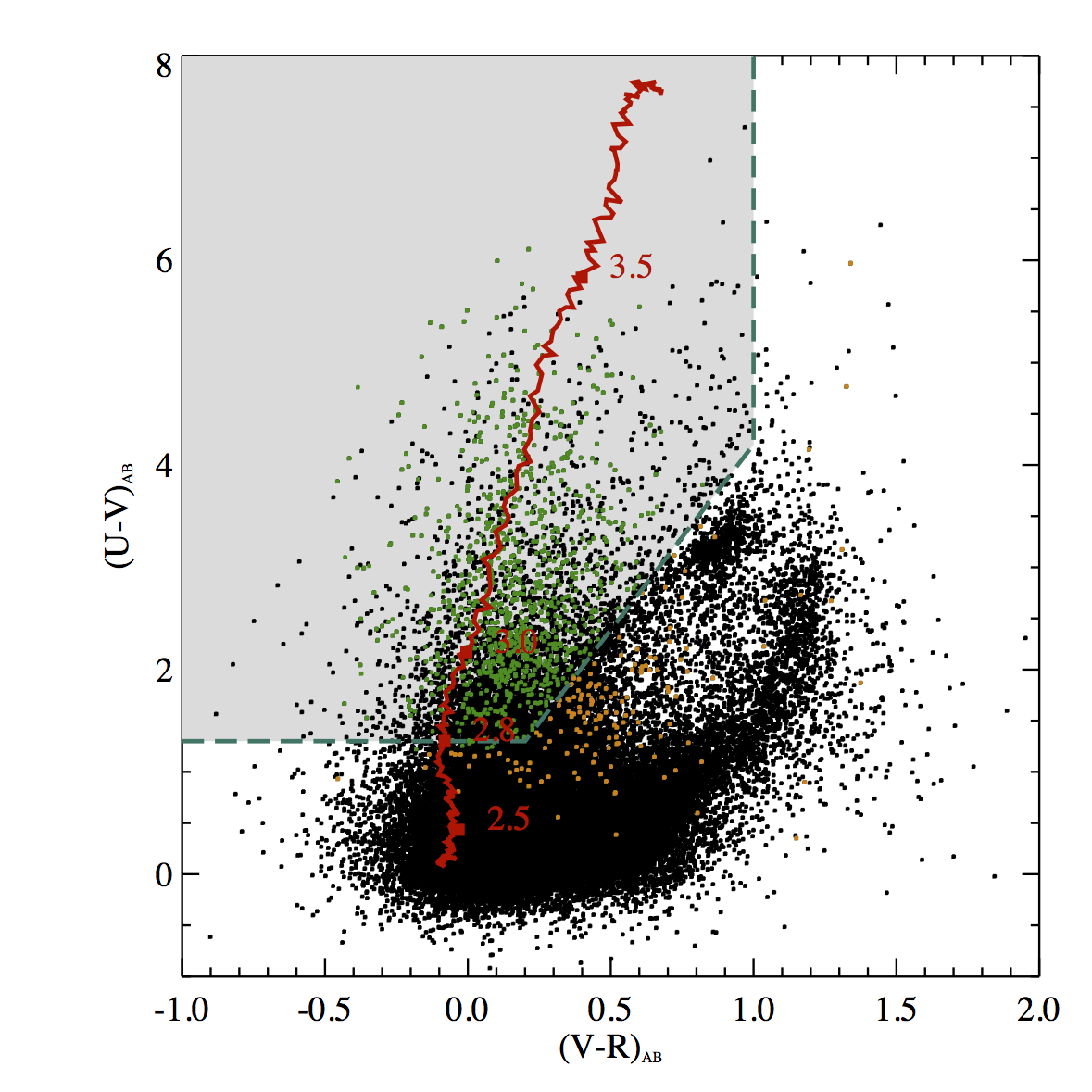}}
\subfigure[   $z\sim$3.6-4.5]{\includegraphics[scale=0.42]{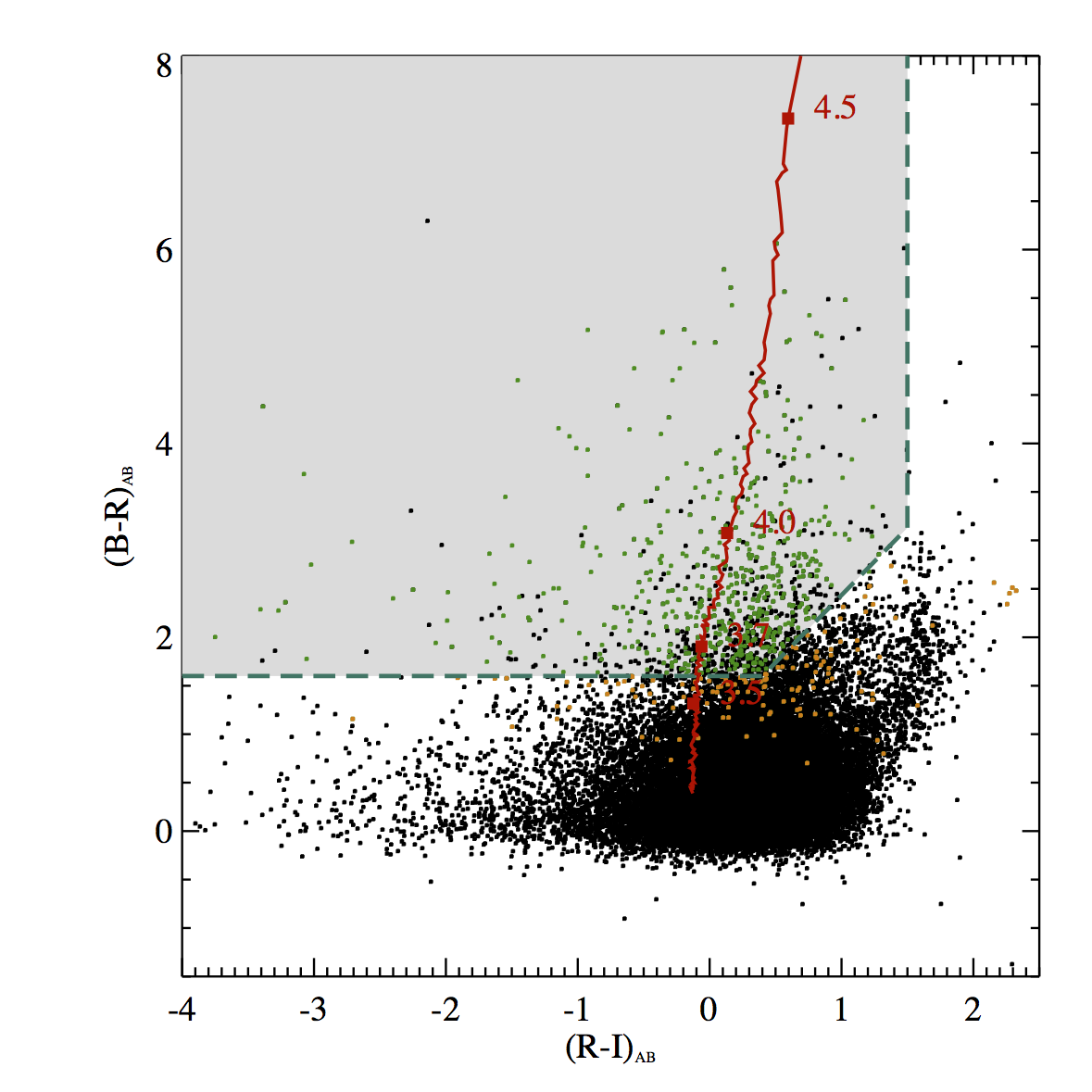}}
\subfigure[   $z\sim$4.7-5.5]{\includegraphics[scale=0.42]{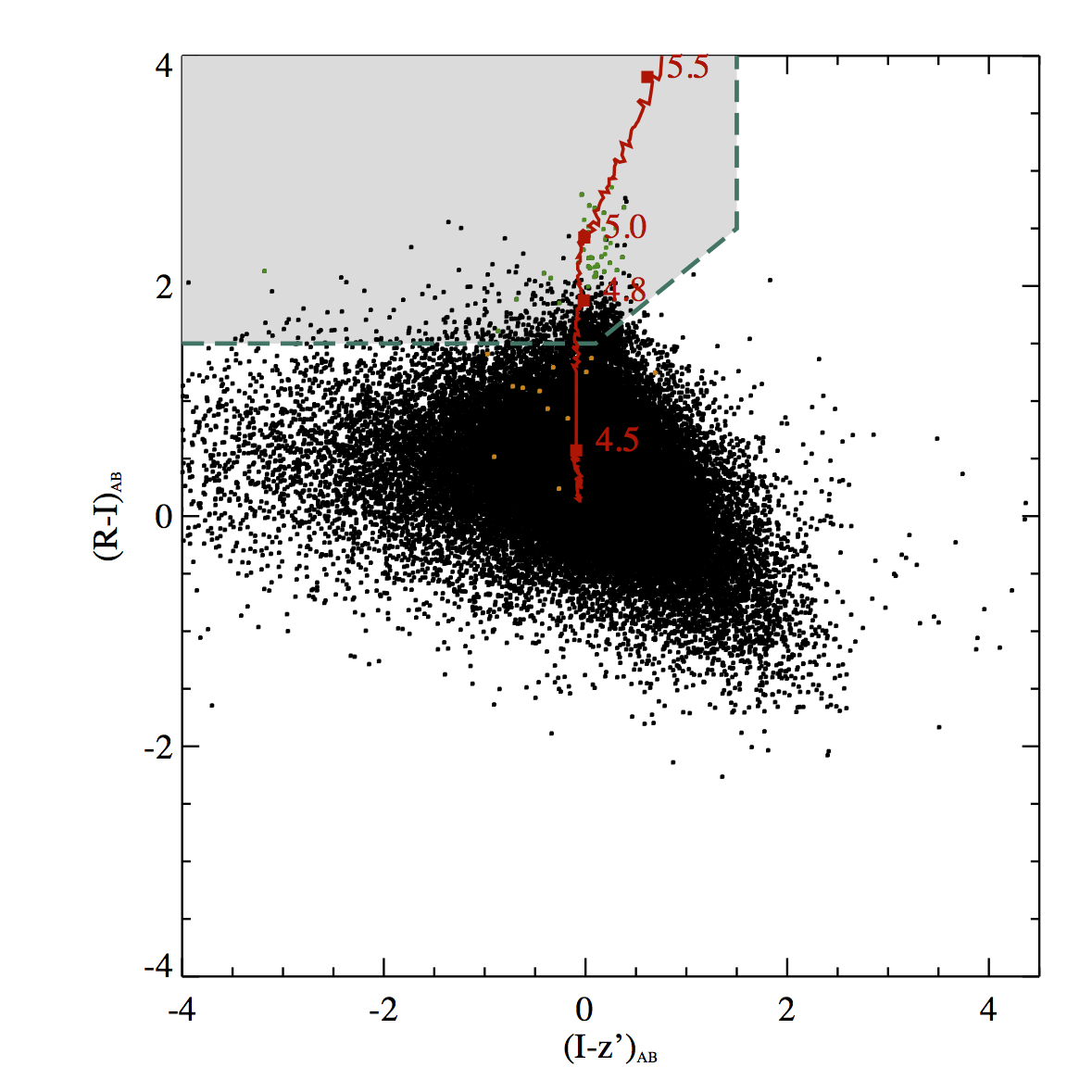}}
\caption{The colour selection of LBGs at $z\sim3, 4$  and 5 in the ECDF-S. The red line represents the colour evolution of a young, metal poor source at each redshift produced using the \citet{Maraston05} stellar population models. All MUSYC sources which meet out magnitude cuts are displayed as black points. Objects which meet our photometric redshift and colour selection criteria are shown as green points (our colour selections are displayed as the grey region bounded by a turquoise dashed line). A small number of sources meet our photometric redshift selection criteria but fall outside of the colour selection window (orange points). Upon inspection of their spectral energy distribution its unclear as to wether or not these sources are truly high redshift star-forming galaxies. Hence, they are omitted from any further analysis.}

\label{fig:col_cuts}
\end{center} 
\end{figure*}

The Large Apex Bolometer Camera \citep[LABOCA, ][]{Siringo09}  is a 295-element bolometer camera mounted on the 12-m Atacama Pathfinder Telescope (APEX) located at Llano de Chajnantor in Chile. It has a bandwidth centre of $\sim$345GHz (870$\mu$m) and half power spectral range of $\sim$60GHz. The pubicly available LESS map comprises 200h of integration time, covering the 30 $\times$ 30\,arcmin$^2$ of the ECDF-S with a uniform coverage of rms\,=\,1.2\,mJy/beam and a resolution of 19\,arcsec (effective beam FWHM). For full details of the LESS survey see \cite{Weiss09}. For our stacking analysis we shall utilise the LESS residual maps, where the 126 sources of $>3.7\sigma$ significance are removed through scaling and subtracting the beam at their positions \cite[see ][]{Weiss09}. We note that we do not stack high redshift galaxies at the positions of these submm bright sources (see below for details).   

The ECDF-S field is extremely well studied, with deep multi-wavelength coverage from X-ray to radio wavelengths. Most notable for this study is the MUSYC survey \citep{Gawiser06}, which provides deep coverage in 32 (broad and intermediate) bands in the optical and near-IR. Such detailed photometric coverage allows the determination of accurate photometric redshifts over a large redshift range \cite[see][for further details]{Cardamone10} .

Initially we produce a sample of photometrically selected LBGs at $z\sim3, 4$ and 5. We apply colour selection criteria to the deep MUSYC catalogue data in order to identify high redshift star-forming galaxies using the Lyman break technique \citep[$e.g.$][]{Steidel95}. These colour selections essentially identify sources with very red colours across the two bluest bands and are aimed at identifying the Lyman break at 1216\AA\ in the rest frame of the galaxy. We also apply additional cuts to reduce contamination from lower redshift sources \cite[see][]{Stanway08}. Colour selections at each redshift are given in Table \ref{tab:col_sel}. In Figure \ref{fig:col_cuts} we display our colour selection criteria designed to identify high redshift star-forming galaxies. We over-plot the redshift colour evolution of a zero extinction, young, metal poor high redshift galaxy produced from the \cite{Maraston05} stellar population models (red line). Any extinction will push source colours to the upper right of the plot. Hence, our colour selections are designed to include both dust-free and extincted sources. We note that using the MUSYC filters, selecting star-forming galaxies at $z<2.8$ is problematic. Hence, we only include sources $2.8<z<3.6$ in our photometrically selected sample at $z\sim3$.

 While these selections identify high redshift star-forming galaxies, they will still be contaminated by low redshift sources. In order to make our sample more robust we use the \cite{Cardamone10} catalog (and references therein) of photometric redshifts in the ECDF-S and only select sources with best fit photometric redshift estimates within each redshift range (see Figure \ref{fig:col_cuts}). We also remove objects whose $\,68\,\%$ confidence (1$\sigma$) error range in redshift extends below the lower boundary of our selection redshifts ($i.e.$ lower than 2.8, 3.6 and 4.5 respectively) in order to remove sources which have poorly-fit photometric redshifts. Therefore, we essentially treat the 32-band photometric redshifts of the \cite{Cardamone10} catalog as a redshift confirmation of our colour-selected sources. We only select sources which both meet our photometric redshift selection and which fall within our colour selection region for high-redshift star-forming galaxies (green points, Figure \ref{fig:col_cuts}).   The requirement of a robust photometric redshift imposes a $R=25.5$ magnitude cut on the sample, as photometric redshifts were only derived for objects brighter than this in the earlier work.

In order to expand our samples to ensure robust statistical results, we also select spectroscopically-confirmed star-forming galaxies at redshift 2.5-3.6, 3.6-4.5 and 4.7-5.5 using the the publicly available redshift surveys in the ECDF-S \citep{Vanzella05, Vanzella06, Vanzella08, Popesso09}. While our photometric sample only extends in redshift down to $z\sim2.8$, we includes spectroscopically confirmed sources down to $z=2.5$ (just 300\,Myr later). Had we used the same  $U-$band filter as that used by \cite{Steidel03}, these objects would have been identified as LBGs (because such a filter has a shorter wavelength red cut-off than that used to select our photometric sample). Given this and the negligible difference in lookback time between $z=2.5$ and $z=2.8$, we are free to include these objects in our lowest redshift sample. Down to $R=25.5$, the magnitude distributions of the photometric and spectroscopic samples are comparable, confirming the spectroscopic sample is a fair representation of the sources selected photometrically.

To constrain this sample to sub-mm faint star-forming galaxies, we then remove galaxies which have (or are spatially close to) sub-mm detections in the LESS map. We define a sub-mm detection as a source which has a $>3\,\sigma$ (870\,$\mu$m) flux in the LESS non-residual map, within 19\,arcsec (the LABOCA effective beam) of its optically derived position - a consequence of this is that no source in our sample is individually detected in the LESS map. This process only removes 33 sources from our samples, with 27 in the $z\sim3$ redshift range. By performing a Monte Carlo analysis of our source positions, we find that we would expect $\sim20$ random correlations of 3$\sigma$ submm bright sources and LBGs at $z\sim3$. Hence, this is consistent with all of these regions being chance superpositions of a LBG and submm source. We also remove any source with a known X-ray counterpart to avoid contamination from high redshift quasars - which show similar colours to star-forming galaxies.

In addition to our LBGs at $2.5<z<5.5$ we also investigate FIR emission from a newly identified sample of spectroscopically confirmed $z\sim4.5$ Lyman-alpha emitters (LAEs) in the same field \citep[][]{Zheng11}. No individual LAE was detected in the LESS map.

Our final robust samples of high redshift star-forming galaxies contain 922, 68 and 20 LBGs at $z\sim3, 4,$ and 5 respectively and 46 LAEs at $z\sim4.5$. We note that our $z\sim5$ sample is comparable to that discussed in \cite{Stanway10} and \cite{Davies10}, and will not provide significantly deeper limits. These sources represent our $full$ high redshift star-forming galaxies samples. However, 
the LBG criteria used in the many studies of $z\gtrsim 3$ galaxies in the past two decades selects objects based on their unobscured UV emission arising from hot young stars. This inevitably identifies a somewhat heterogeneous sample of sources over a range of stellar mass and dust extinction. A  comparatively low mass source can have the same observed UV properties as a more massive system with a higher extinction and higher overall star formation rate (and with a higher FIR luminosity).  So while it is valid to explore the FIR properties of sources that are selected using standard LBG criteria, precisely because they represent a type of system that has the basis for a multitude of previous studies of $z\gtrsim3$ galaxies, it is informative to compare these to other samples selected not just as LBGs, but with constraints on their stellar masses and extinction-corrected star formation rates.  Consequently in this work we derive and explore several subsamples based on these criteria, restricting these to objects at $z\sim 3$ as at this redshift there is sufficient data to make this comparison between the subsamples and the main LBG sample. We describe each of these subsamples in the following section.

\begin{table*}
\centering

\begin{scriptsize}

\begin{tabular}{c c c c c c }
\hline
\hline
Redshift & Colour & Colour & Colour & Magnitude & IR colour\\
\hline

2.8-3.6 & U - V $ > $ 1.2 & V - R $ < $ 1.1 & U - V $ > $ 3.63(V - R) + 0.58&V $ < $ 26.2 & - \\
3.6-4.5 & B - R $ > $ 1.6 & R - I $ < $ 1.5 & B - R $ > $ 1.27(R - I) + 1.1&R $ < $ 27.0& I - J $ < $ 1.0\\
4.5-5.5 & R - I $ > $ 1.5 & I - z $ < $ 1.4& R - I $ > $ 0.78(I - z) + 1.7& 23.0 $< $ I  $< $ 25.8  & I - J $ < $ 1.0   \\
 
\hline
\end{tabular}
\end{scriptsize}
\caption{The colour selection criteria used to identify star-forming galaxies at $z\sim3, 4$ and 5. An additional I-J colour selection is applied at $z\sim4$ and 5 to reduce contamination from low redshift sources \citep[see][]{Stanway08}. }

\label{tab:col_sel}        
\end{table*}

\subsection{Subsamples of the most massive, most actively star-forming and UV bright rest-frame UV/optical-selected systems at $z\sim3$}
\label{sec:high_SFR}

\subsubsection{Stellar-mass selected subsamples}
 
Previous work, \citep[e.g.][]{magdis101}, hereafter M10 has indicated that at $z\sim 3$ the flux in the observed Spitzer/IRAC bands (rest-frame near-IR) is directly related to the stellar mass of an LBG. In order to select LBGs down to a given stellar mass, we follow M10. We select one subsample with all sources  detected at M\,$<\,25.0$ in at least one of the IRAC bands \citep[comparable to the M10 sample - IRAC magnitudes are taken from the ][catalogue]{Cardamone10} and  a second brighter subsample  at  M\,$<\,22.5$ (10 times more massive assuming a direct relationship between IRAC magnitude and stellar mass).  We then also exclude sources which have an X-ray detection in the deep Chandra maps of the field (to rule out bright AGN), and once again remove sources which are within 19$^{\prime\prime}$ of a $>3\sigma$ source in the LESS map. Following M10, we use only the spectroscopically-confirmed LBGs at $z\sim 3$. The fainter subsample (IRAC-25 hereafter) selects almost all the spectroscopically-selected LBGs (405 of 425 sources, compared to 51 sources in M10), while the brighter sample (IRAC-22.5) consists of  50 galaxies.  All values given in the tables of this paper are for the brighter  sample (see Section \ref{sec:IRAC_c} for further details).        
 
M10 use the deep AzTEC 1.1mm observations of the GOODS-N field to produce  a composite image of the positions of their IRAC-detected LBGs. This composite image displays a 0.41\,mJy ($\sim3.7\sigma$) detection of their sample at 1.1mm. This suggests that, while typical $z\sim3$ LBGs remain undetected at sub-mm wavelengths, a stacking analysis of the most massive LBGs should yield a detection. Using our stellar mass selected samples we shall investigate this further. Assuming a dust SED with power law emissivity, $\beta_{FIR}$=2, the M10 work predicts an 870$\mu$m flux of $\sim0.65\,$mJy for a similar sample of IRAC$<$25.0 $z\sim3$ LBGs.

While the M10 result may suggest that the most massive $z\sim3$ LBGs are the most FIR bright (hence, there is a correlation between stellar mass and FIR emission), it is unclear as to whether or not large stellar masses are the prerequisite of observable FIR fluxes. The galaxies in their sample display z - 3.6$\mu m$ colours indicative of an old stellar population. In addition, these sources have reasonably large UV luminosities, indicating that they are undergoing a significant burst of unobscured star-formation. Hence, either age, unobscured SFR or both may also be a significant factor in determining whether or not a LBG is detectable at FIR wavelengths.

To investigate this further we also produce two subsamples of $z\sim3$ galaxies, one with the largest extinction corrected star-formation rates \citep[$e.g.$][]{Adelberger00} and another with the largest UV fluxes \citep[representing a typical $z\sim3$ Lyman-break selected sample $e.g.$][]{Steidel95}. 

\subsubsection{ Extinction-corrected high SFR sample}

In order to select a subsample with the largest extinction corrected star-formation rates we apply an extinction correction to the observed rest-frame UV magnitude following the same procedure as that outlined in \cite{Adelberger00}. We estimate the UV spectral slope ($\beta_{UV}$) using the best fit to the MUSYC broad and intermediate band observations in the 5000-9000\AA\ (rest-frame $\sim1250-2250$\AA) range. We then calculate a rest-frame UV extinction at 1600\AA\ ($A_{1600\AA}$) using:

\begin{equation}
A_{1600\AA}=4.43+1.99\,\beta_{UV}
\label{eq:ex_cor}
\end{equation}

We apply this extinction correction to our spectroscopically confirmed sample, k-correct and calculate extinction corrected star-formation rates at rest-frame 1600\AA\ in all sources, using the standard conversion between rest-frame UV flux and SFR \cite[irrespective of $\beta_{UV}$, $e.g.$][]{Rosa-g02}:

 \begin{equation}
\mathrm{SFR_{UV}\, (M_{\odot}/yr)=1.4 \times 10^{-28}\,L_{UV} \, (ergs/sec/Hz)} 
\label{eq:UV_SFR}
\end{equation}  

We then select the top 50, 100, 150, 200 and 250 most highly star-forming sources at $z\sim3$. We find that the top 200 sources (equating to an extinction corrected SFR$>$6.7M$_{\odot}$yr$^{-1}$) produces the highest signal to noise in our resultant analysis - hence, we shall only discuss this sample (hereafter the high-pSFR sample). This extinction correction assumes that each galaxy has an intrinsic (unattenuated) $\beta_{UV} \sim-2$ and relies on this premise to predict total star-formation rates \citep[an intrinsic $\beta_{UV} \sim-2$ model is found to be appropriate for both high redshift LBGs and local analogues, $e.g.$][]{Overzier11}. Therefore, this sample is essentially those sources with the highest $predicted$ SFRs, with the caveat of an assumed $\beta_{UV}=-2$, with any deviation from this slope is caused by dust attenuation. Following the prescription of \cite{Meurer99}, this implies that sources with the largest $predicted$ SFRs should display the largest FIR fluxes. Here, we can directly test this prediction by comparing the FIR emission inferred by the assumption that all sources have an intrinsic $\beta_{UV}=-2$, and those directly measured in the FIR. The sample is clearly dominated by those sources with UV spectral slopes that deviate most strongly from $\beta_{UV}=-2$. If, for these sources, the dominant cause of this deviation is not dust extinction, but something else, e.g. variation in stellar population, the sources may be less luminous in the FIR than predicted.  As we shall see in Section \ref{sec:uv_slope}, this appears to be the case - suggesting that the unattenuated UV-spectral slope is unlikely to be -2 (in these sources). Therefore while we label this sample as `high-pSFR', we note that this is only true by the standards of previous work which have assumed a constant intrinsic UV spectral slope, and later we shall show that this is unlikely to be the case.                 
  
\subsubsection{UV-brightest subsample}

The UV-brightest subsample, which represents the LBGs with the highest unobscured star formation rate, is simply a subset of our full sample cut at a higher $R-$band magnitude ($R=24.43$ rather than $R=25.5$), the limit chosen to select 200 sources.  The choice of sample size is once again assigned retrospectively to achieve the maximum signal to noise in our resultant stacked images (see below), while restricting the sample to the most rest-frame UV-bright sources. This sample contains the UV-bright end (20-25\%) of our full LBG selection at $z\sim3$. The sources are not particularly extreme or rare examples of LBGs;  they represent those  which are often the subject of  spectroscopic studies of $\sim 3$ LBGs  \citep[$e.g.$][]{Shapley01}. 
 
Using our samples of the most massive (IRAC-25, IRAC-22.5), most actively star-forming (high-pSFR, with the caveats discussed above) and rest-frame UV-brightest (UV-bright) systems we investigate potential correlations between stellar mass, star-formation rates and FIR flux at $z\sim3$. We estimate the mean stellar mass of each sample using the method outlined in \cite{Magdis_c}, which relates IRAC 8.0$\mu$m magnitude to stellar mass:

 \begin{equation}
\mathrm{Log(M_{*}/M_{\odot})=2.01-0.35M_{8.0}} 
\label{eq:mass}
\end{equation}

where $M_{8.0}$ represents the mean 8.0$\mu$m magnitude of each sample. The UV-optical properties of all $z\sim3$ samples are given in Table \ref{tab:UV_props}.

 \begin{table*}
\centering

\begin{scriptsize}
\begin{tabular}{c c c c c c c c c }
\hline
\hline
Sample & Selection & N$_{gals}$ & M$_{*}^{1}$  & $\beta_{UV}$ & SFR$_{UVobs}^{3}$ & SFR$_{UVex}^{3}$ & Predicted SFR$_{FIR}^{4}$ & Predicted L$_{FIR}^{5}$ \\
 & & & (Log[M$_{\odot}$])& &(M$_{\odot}$\,yr$^{-1}$) &(M$_{\odot}$\,yr$^{-1}$) &(M$_{\odot}$\,yr$^{-1}$) &(Log[L$_{\odot}$]) \\ 
\hline
IRAC-22.5 & spec, 3.6$\mu$m $<$ 22.5 & 50  & 10.7 & -1.3 & 33.9 & 204.5 & 182 & 12.0 \\

high-pSFR & spec, SFR$_{UV-ex} > 6.7$ & 200 & 10.4 & -1.4 & 24.1 & 148.2 & 124 & 11.9 \\

UV-bright & u-drop \& spec, $R <$ 24.43 & 200 & 10.4 & -1.8 & 24.9 & 75.6 & 52 & 11.5 \\

All $z\sim3$ & u-drop \& spec & 922 & 10.2 & -1.8 & 7.7 &  55.4 & 48 & 11.5 \\

\hline
\end{tabular}
\end{scriptsize}

\caption{The UV-optical properties of $z\sim3$ samples used in this work. The IRAC-22.5 and high-pSFR samples are only drawn from spectroscopically confirmed $z\sim3$ sources only, while the UV-bright and all $z\sim3$ samples contain both spectroscopically confirmed galaxies and photometrically selected sources. $^{1}$Value derived from mean 8.0$\mu$m flux, all other values are mean values taken from all sources in each sample. $^{2}$ Observed UV SFR calculated using equation \ref{eq:UV_SFR} and the source luminosity at 1600\AA . $^{3}$ Extinction corrected SFR calculated from the luminosity at 1600\AA\ with an extinction correction applied (equation \ref{eq:ex_cor}). $^{4}$ Predicted FIR SFR taken as the extinction corrected SFR - observed UV SFR. $^{5}$ Predicted FIR luminosity derived from the predicted FIR SFR.}
\label{tab:UV_props}
\end{table*}

\section{Analysis}
\label{sec:anal}

In order to constrain the FIR emission from the average source, data were combined from the positions of all sources in each of our samples. A 30$\times$30 pixel ($\sim10 \times 10$ beam size) region centred on each source was extracted from the LESS residual map. These 30$\times$30 pixel regions were then combined into an average image taking the mean flux at each pixel position over all of the extracted regions (stacking). However, stacking processes involving large-beam deep sub-mm maps which are close to the confusion limit (such as LESS) can be biased by flux arising from both faint and bright nearby sources \citep[see][for more detailed discussion]{Lutz10}. Firstly, the number-density of faint background sources at the flux level of the LESS map ($\sim$ 1.2\,mJy) is comparable to the LABOCA beam density. Therefore, there is essentially no blank sky background and all LABOCA beam positions contain flux arising from faint background objects. Secondly, any LABOCA beam position may contain additional flux from the wings of the beam of a nearby bright source. \cite{Lutz10} perform simple Monte Carlo simulations of model LABOCA data in order to estimate the flux contribution to LESS beam positions from both faint background sources and nearby bright sources. They find that the LESS residual map requires the subtraction of 0.072\,mJy\,beam$^{-1}$ in order to remove the contribution from such sources. Hence, any detection or limit obtained in our stacking analysis must be scaled to reflect this additional flux. Therefore, prior to stacking we subtract 0.072\,mJy\,beam$^{-1}$ from the LESS residual maps, allowing any detected flux in our stacked image to be used directly. However, if no emission is detected in the stacked images, the measured RMS will essentially measure fluctuations about this scaling factor. Hence, any further analysis must use a limit of 0.072(+rms)\,mJy\,beam$^{-1}$.

\section{Results for the $z\sim 3$ samples}

\subsection{The IRAC-25  and IRAC-22.5 samples} 
\label{sec:IRAC_c}

\begin{figure*}
\begin{center}

\includegraphics[scale=0.35]{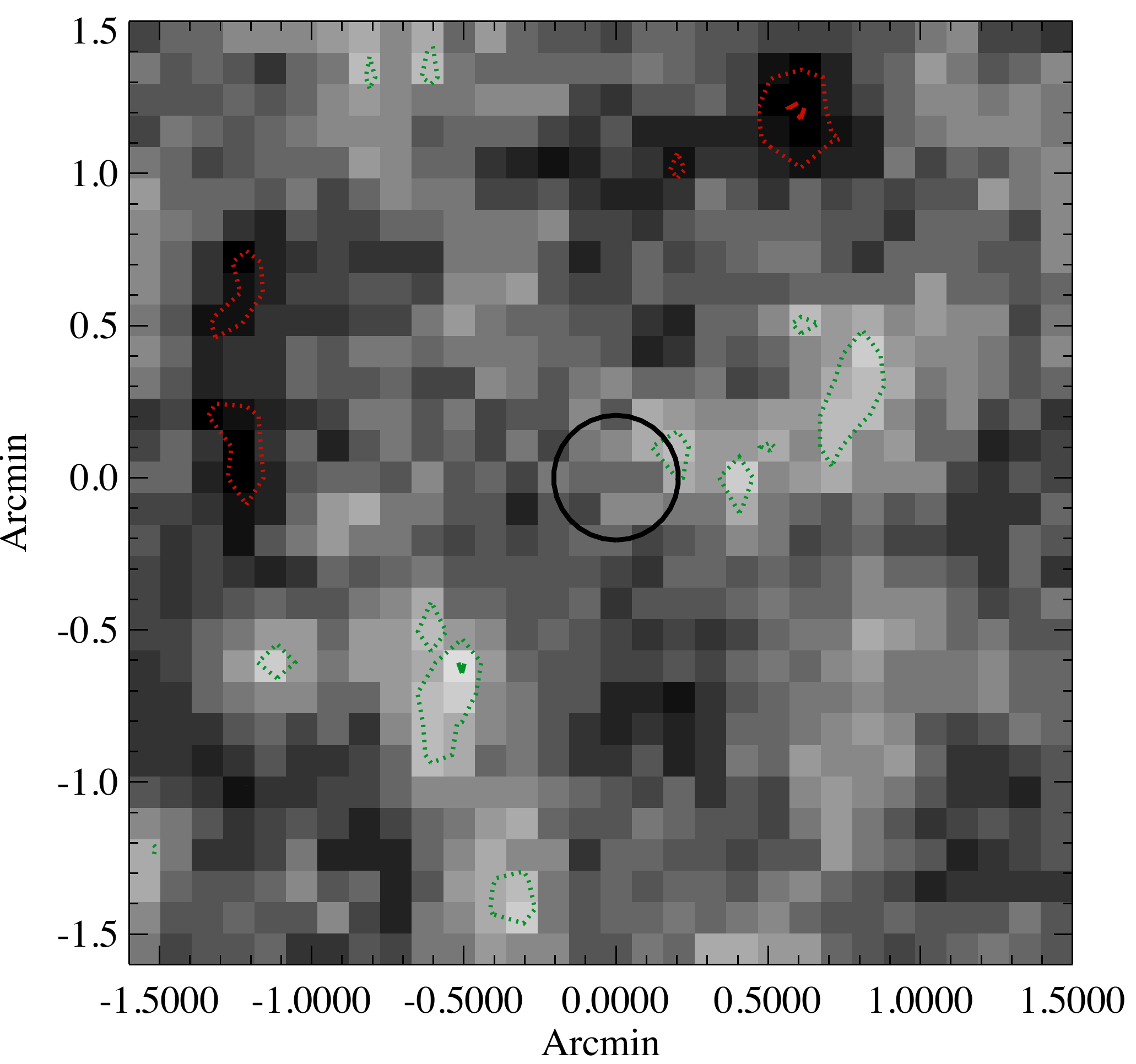}
\includegraphics[scale=0.35]{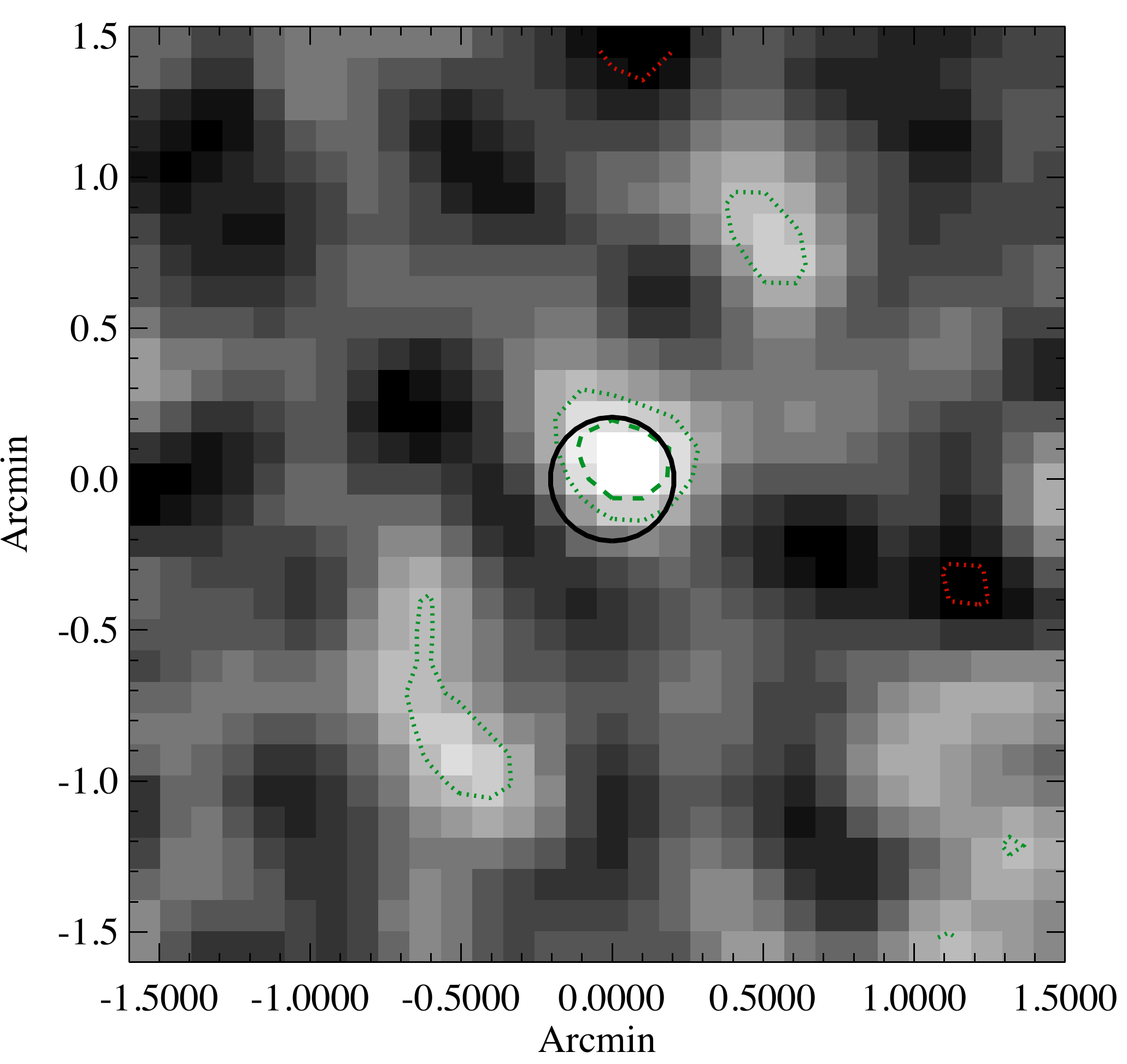}\\
\includegraphics[scale=0.35]{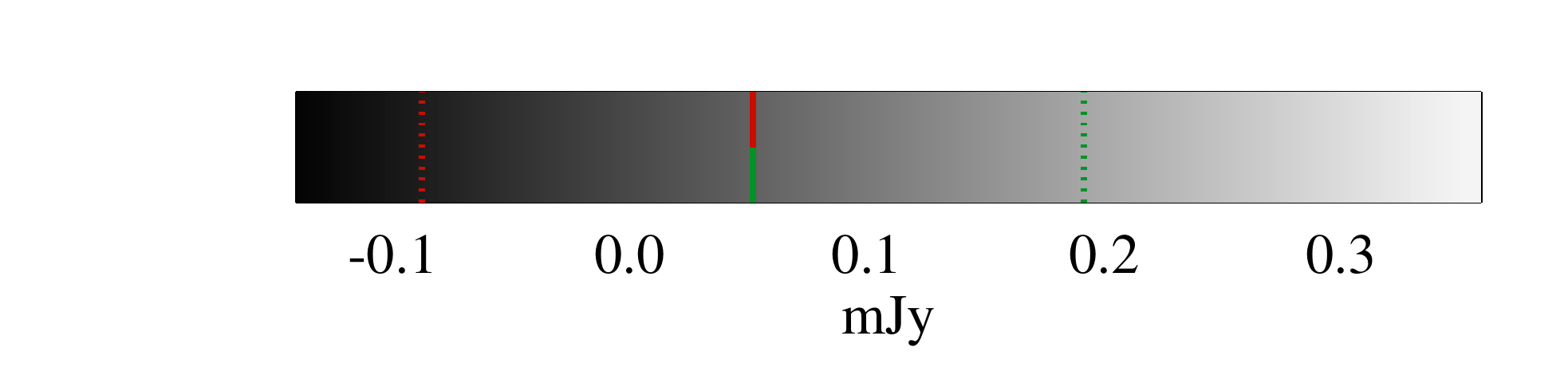}
\includegraphics[scale=0.35]{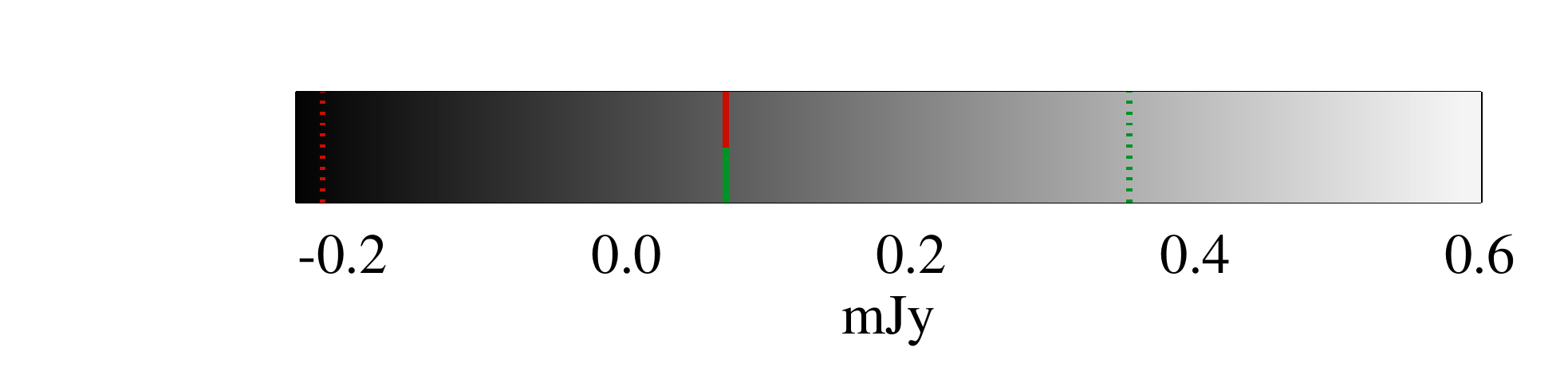}\\

\includegraphics[scale=0.35]{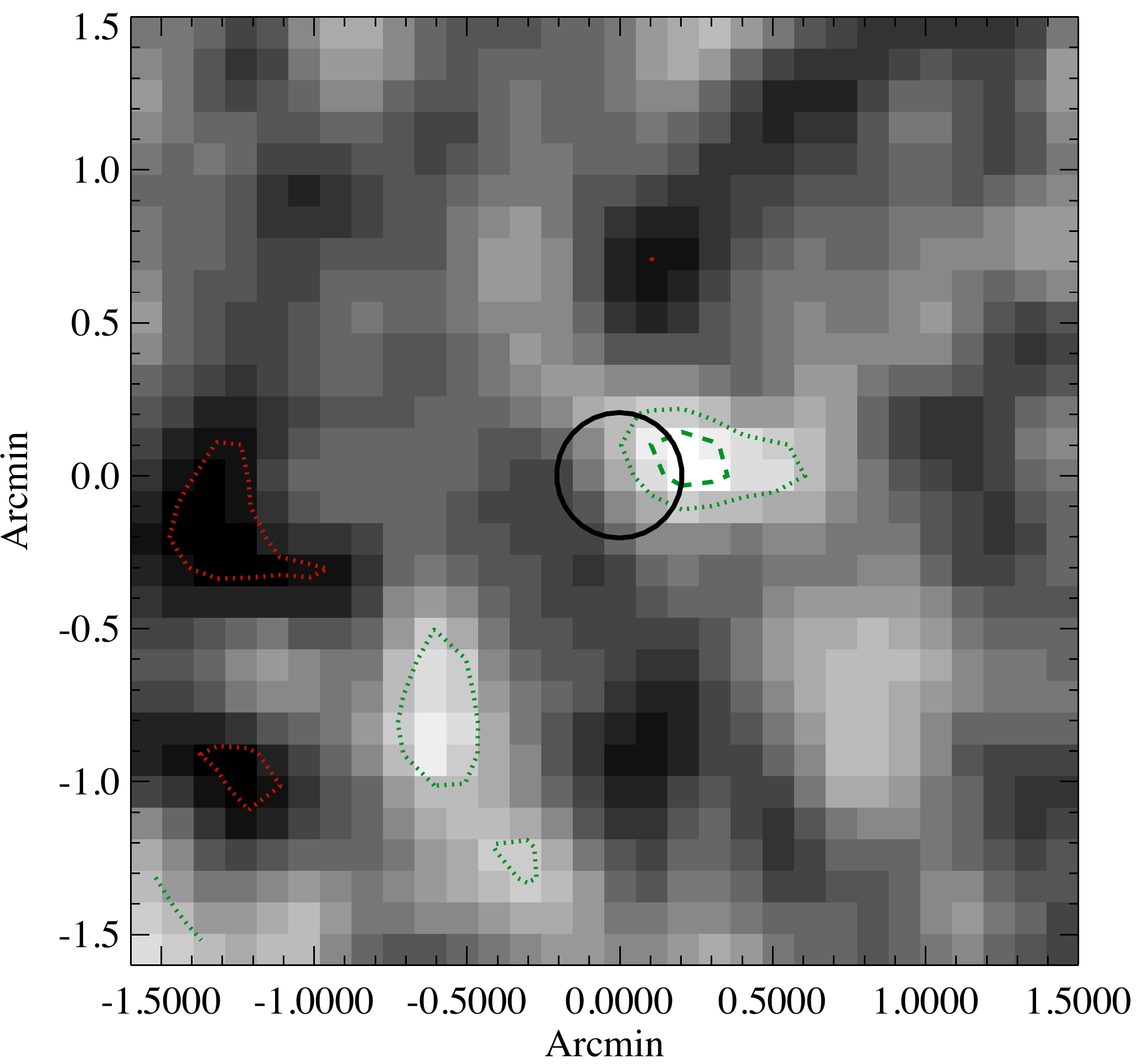}
\includegraphics[scale=0.35]{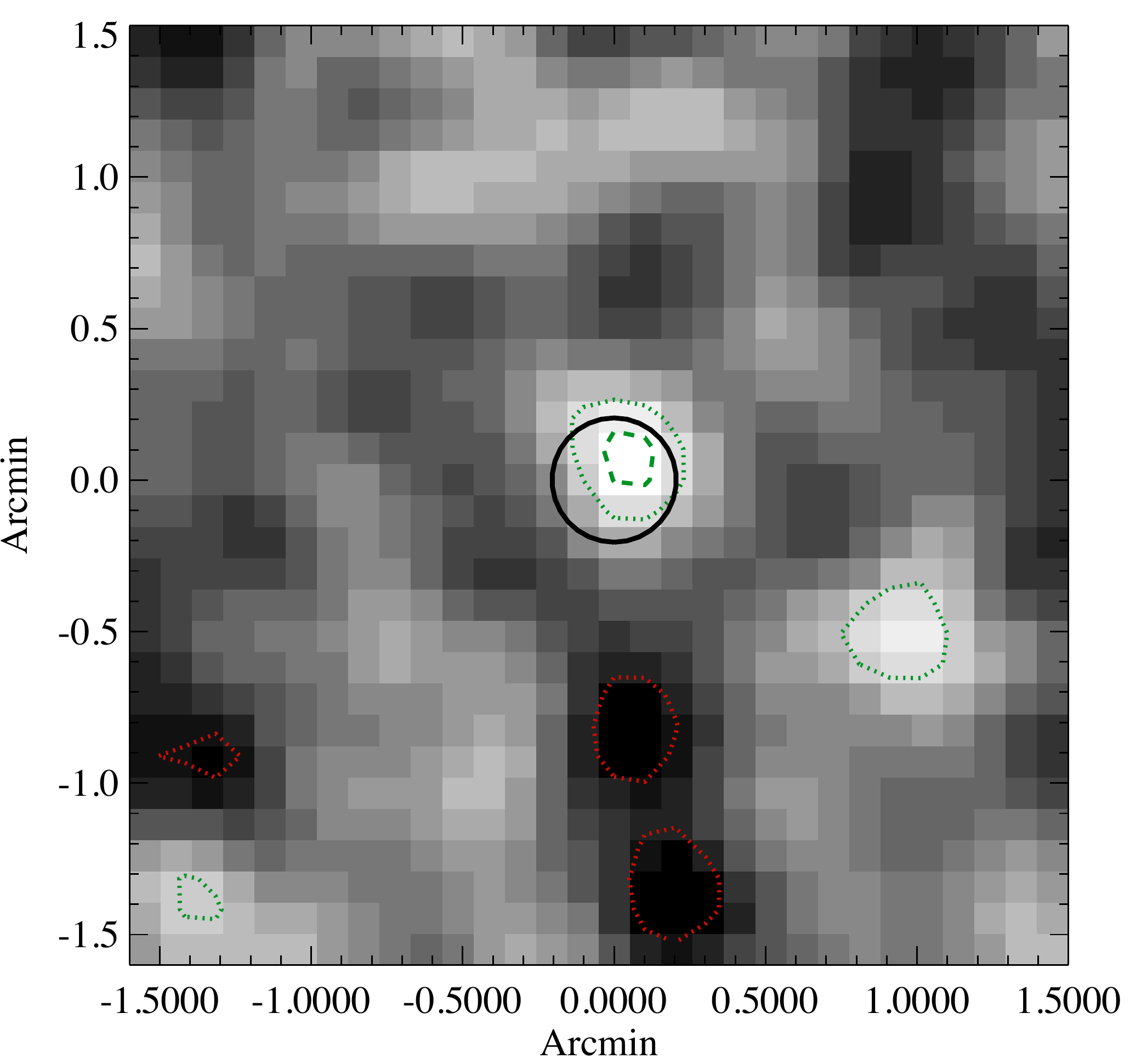}\\
\includegraphics[scale=0.35]{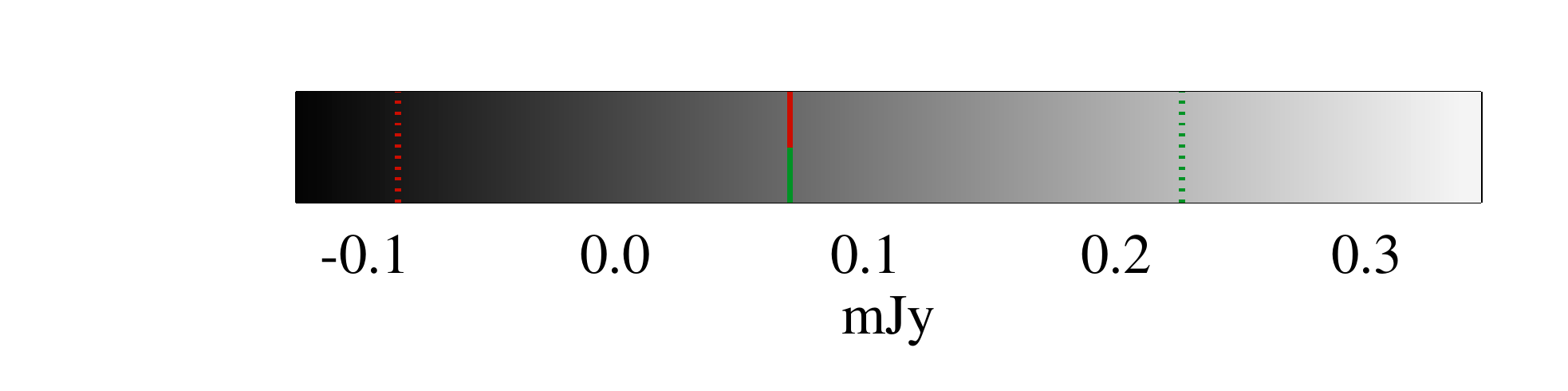}
\includegraphics[scale=0.35]{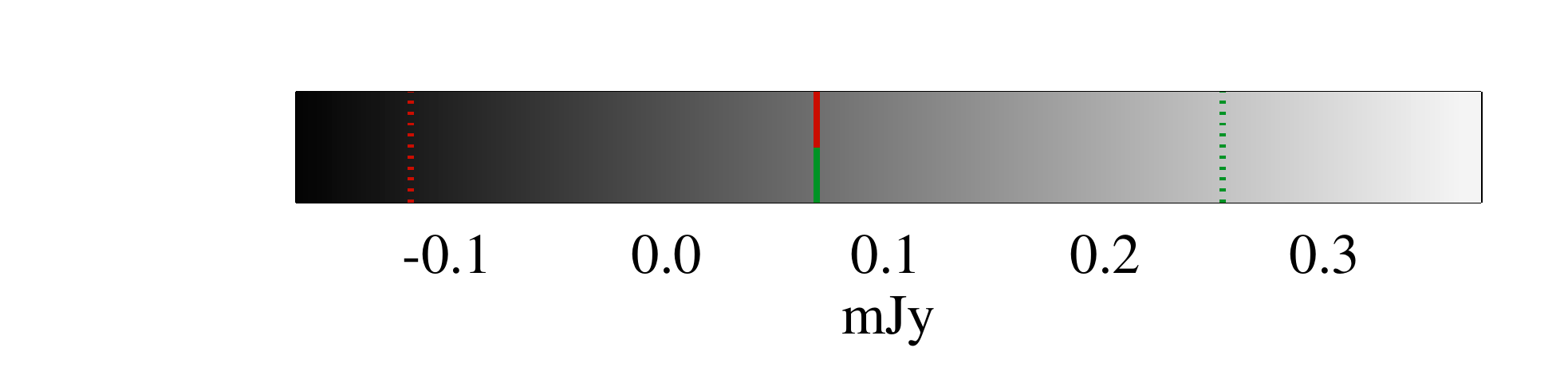}\\

\caption{ Composite 870$\mu$m image of the positions of our subsamples convolved with a gaussian of the same FWHM as the LABOCA beam size: \textbf{Top left:} The IRAC-25, spectroscopically confirmed $z\sim3$ sources. This sample represents a comparable (but $\sim8 \times$ larger) set of sources to the M10 sample of LBGs in the GOODS-N field, which obtained a detection of the typical source at 1.1mm at a 0.41\,mJy level. We do not detect the typical $z\sim3$ IRAC-detected source in the ECDF-S at a $\sim0.2$mJy, 2 $\times$ rms level. \textbf{Top right:} The further constrained IRAC $<$ 22.5, and hence most massive, spectroscopically confirmed star-forming sources at $z\sim3$ (the IRAC-22.5 sample), displaying a detection at $\sim0.61$mJy (3.9 $\times$ rms). \textbf{Bottom left:} The high-pSFR sample (SFR$_{UV-ex}>$ 6.7M$_{\odot}$yr$^{-1}$), where we obtain a detection at a $\sim0.35$mJy, 4.3 $\times$ rms level. \textbf{Bottom right:} The sample of $z\sim3 R <$ 24.43 (and hence most rest-frame UV-bright) LBGs, where we obtain a detection at a $\sim0.36$mJy, 3.9 $\times$ rms level. In all figures contours display both positive (green) and negative (red) deviations of 2 (dotted line) and 3 (dashed line) $\times$ rms away from the mean value in the field. The LABOCA beam size is displayed as the black circle. Note that the colour scaling is not consistent over all images.}

\label{fig:IRAC_d}
\end{center} 
\end{figure*}

Figure \ref{fig:IRAC_d} (top left) displays the composite image produced from the IRAC-25 sample. We find that the stacked source is undetected at a $\sim0.2$\,mJy (2 $\times$ rms) level. This is surprising given the M10 detection in a comparable sample at $\sim0.4$mJy at 1.1mm (predicted 0.65\,mJy at 870$\mu$m). In order to validate our stacking procedure, we apply our method to a previously published stacked detection in the same field. \cite{Greve10} use the LESS survey to produce a composite image of $\sim600$ star-forming $BzK$ ($sBzK$) selected galaxies in the ECDF-S, obtaining a $\sim0.4$\,mJy detection of the average $sBzK$ in the field. We apply our stacking procedure to an identical sample of $sBzK$ sources in the field and also obtain a $\sim$0.4\,mJy detection - thus suggesting our method is sound.  In an additional test we repeat the procedure outlined in M10. We select their sample of spectroscopically confirmed LBGs in the GOODS-N and apply similar cuts to those described in their work. We then combine the publicly-available AzTEC data at their source positions using three independently produced stacking procedures and still do not obtain a detection.

Carrying out the same stacking procedure on the brighter IRAC-22.5 sample (which contains the brightest 12 per cent of the sources in the fainter sample), leads to a $4\times$ rms) detection. The top right panel of Figure \ref{fig:IRAC_d} shows the composite image of this sample displaying a 0.61$\pm0.14$\,mJy source at the expected position. Clearly if the fainter sample had the same ratio of IRAC to 870$\mu$m flux as the bright sample, we would not have detected it.  Further discussion of stellar-mass selected samples is  limited to the brighter, IRAC-22.5 sample.

\subsubsection{FIR SED of the IRAC-22.5 sample}

\begin{figure*}
\begin{center}

\includegraphics[scale=0.2]{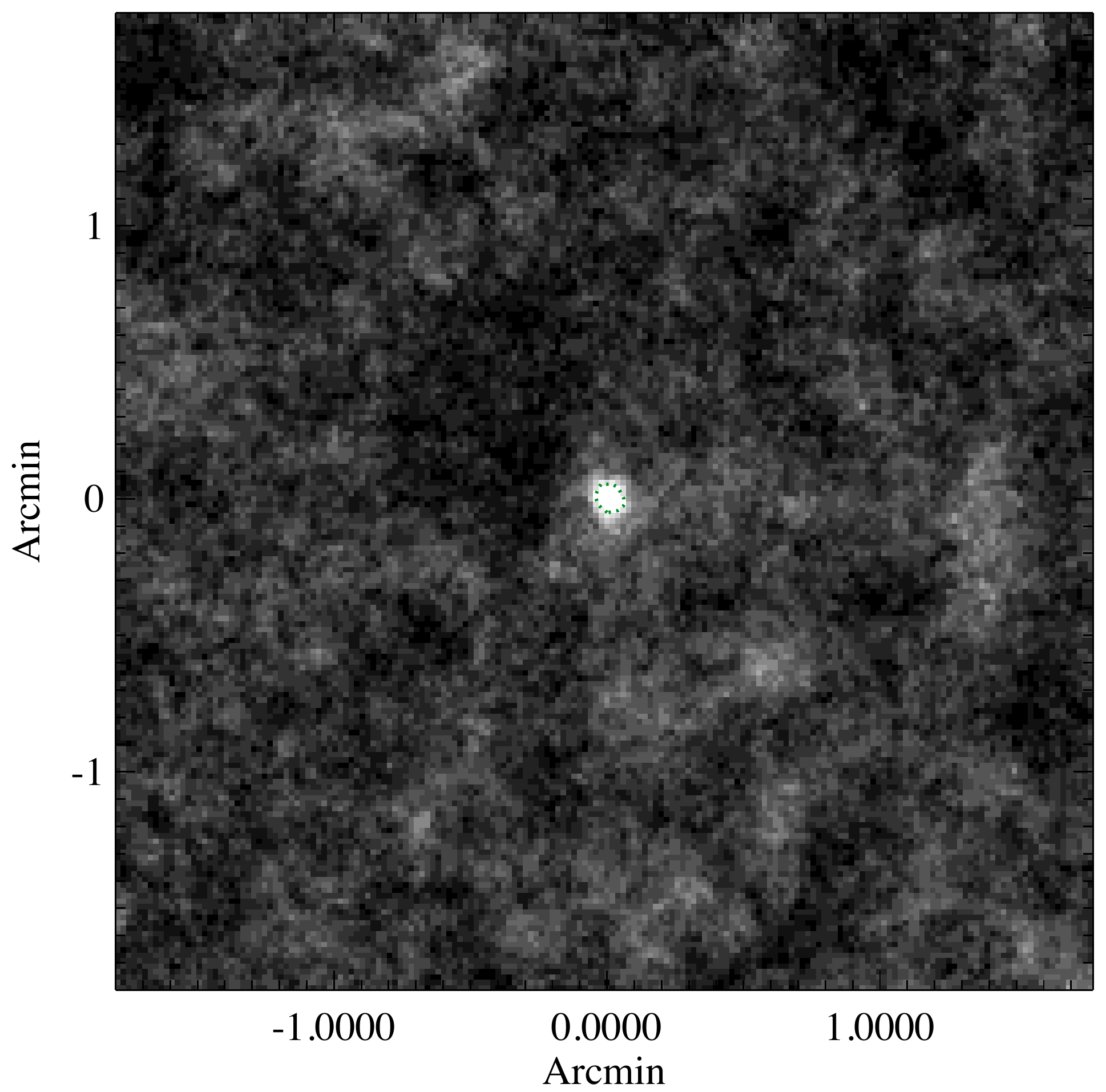}
\includegraphics[scale=0.2]{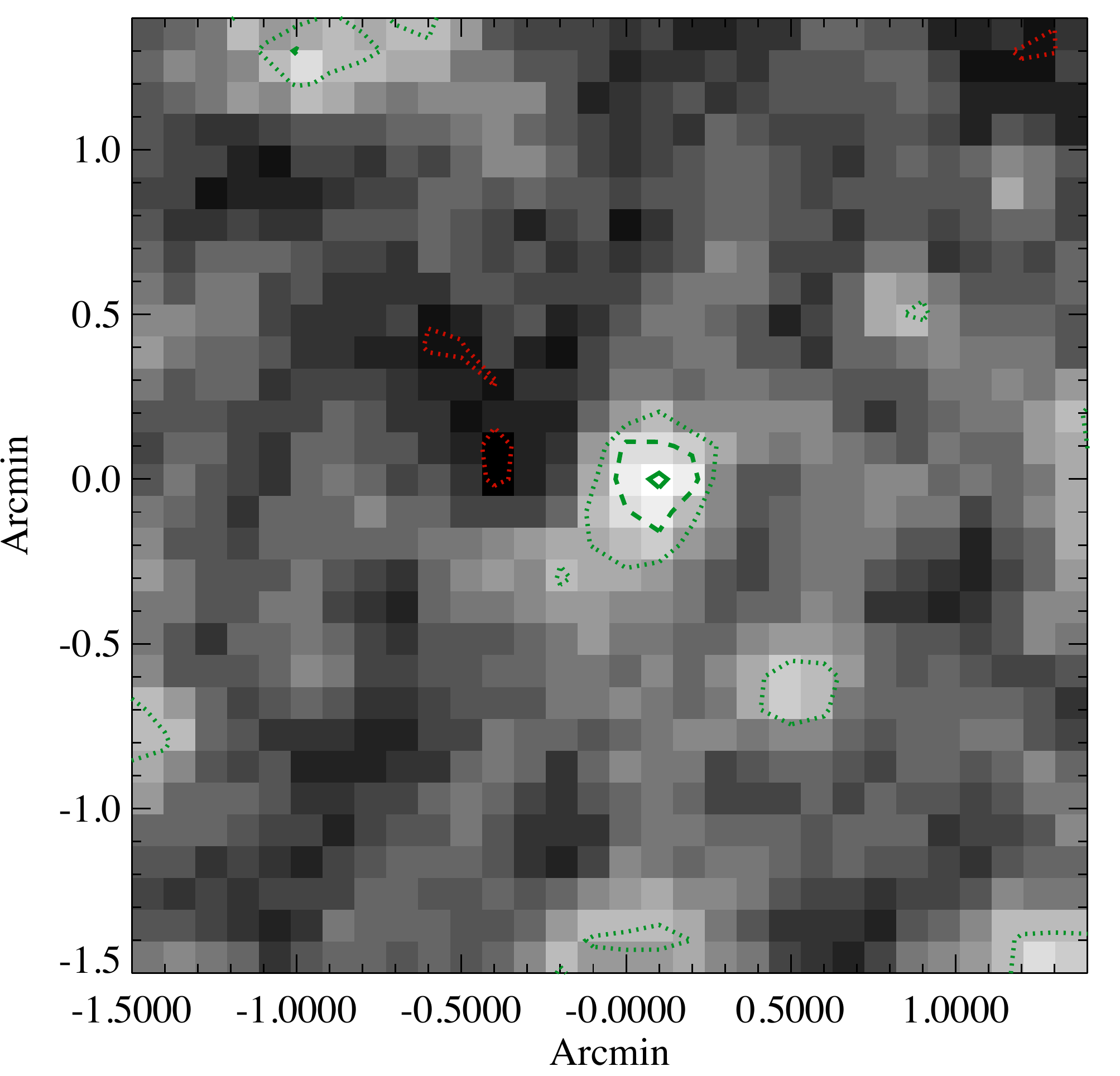}
\includegraphics[scale=0.2]{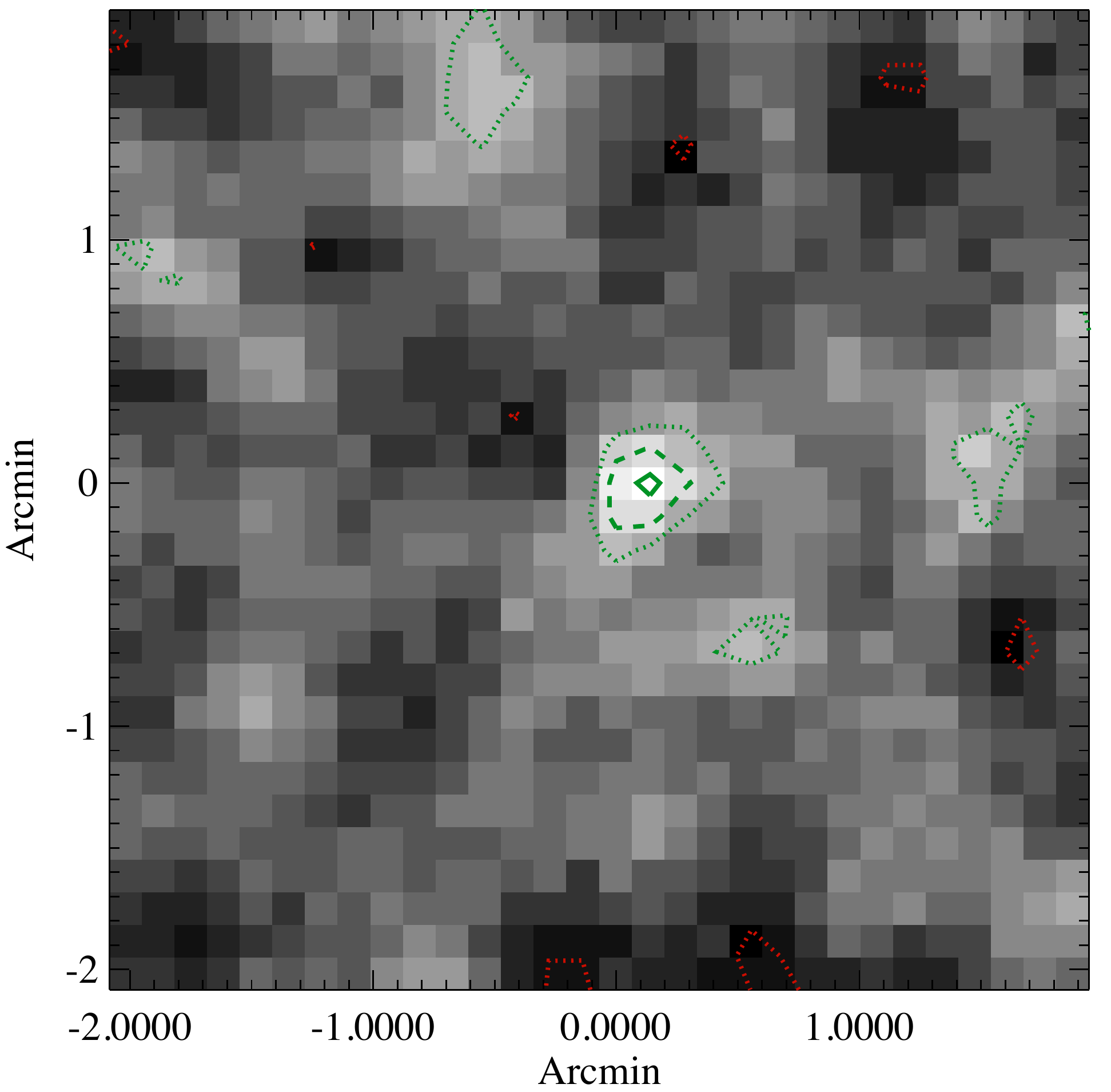}
\includegraphics[scale=0.2]{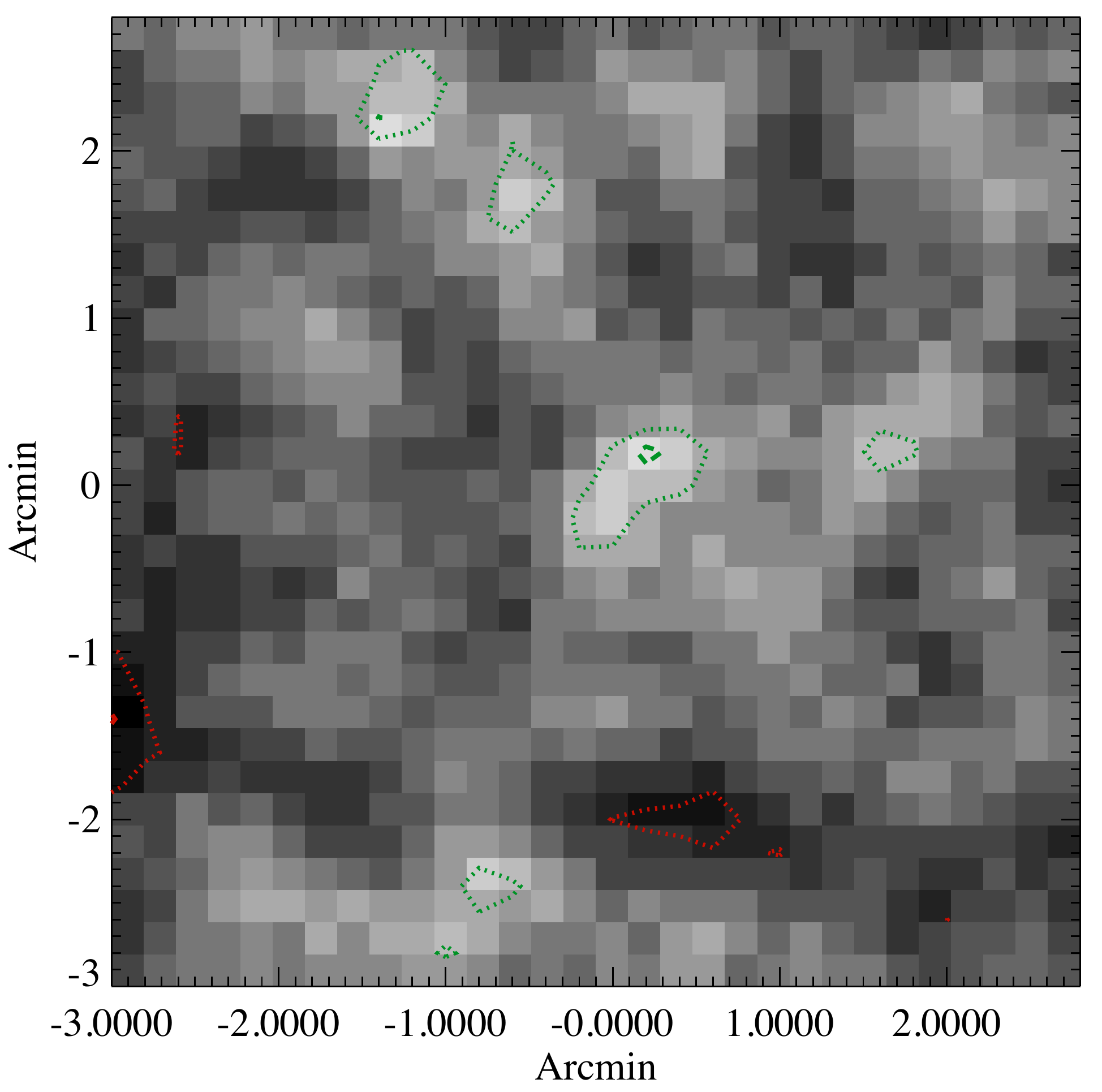}\\

\includegraphics[scale=0.2]{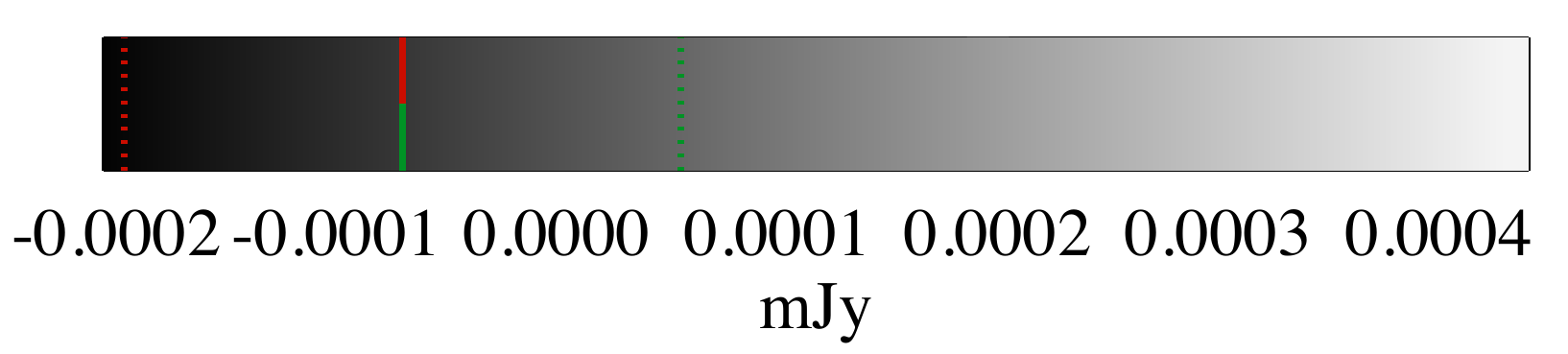}
\includegraphics[scale=0.2]{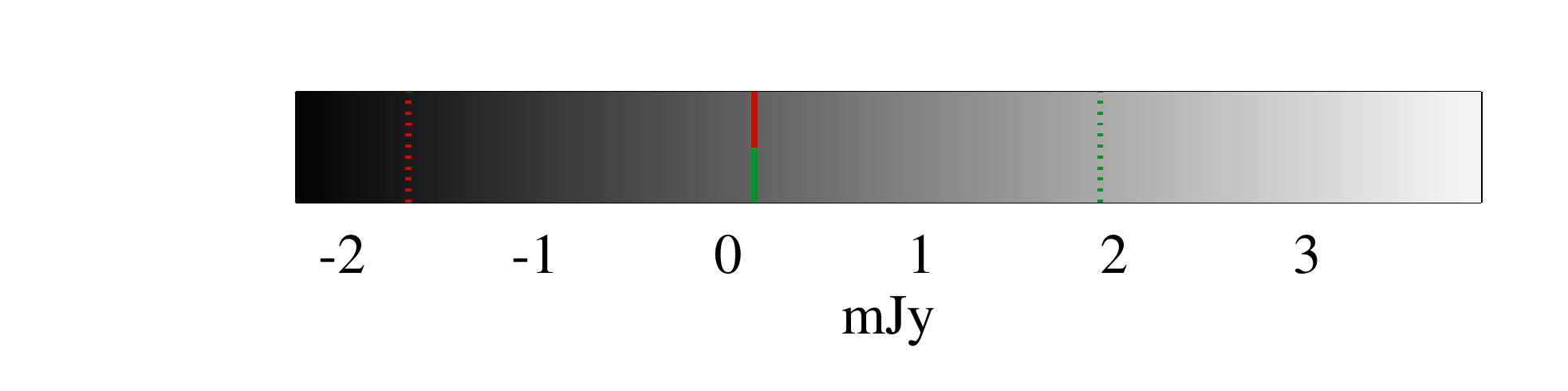}
\includegraphics[scale=0.2]{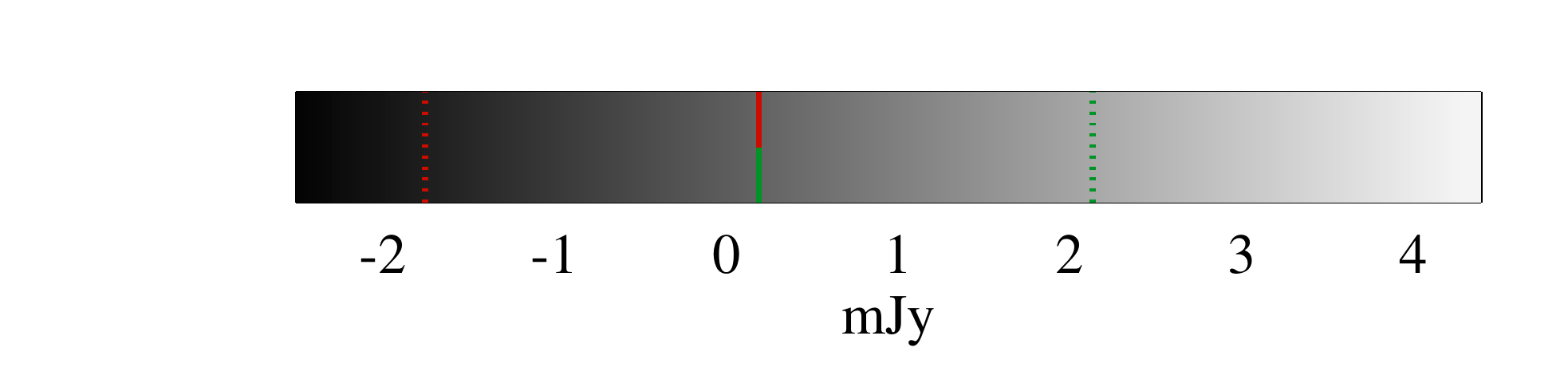}
\includegraphics[scale=0.2]{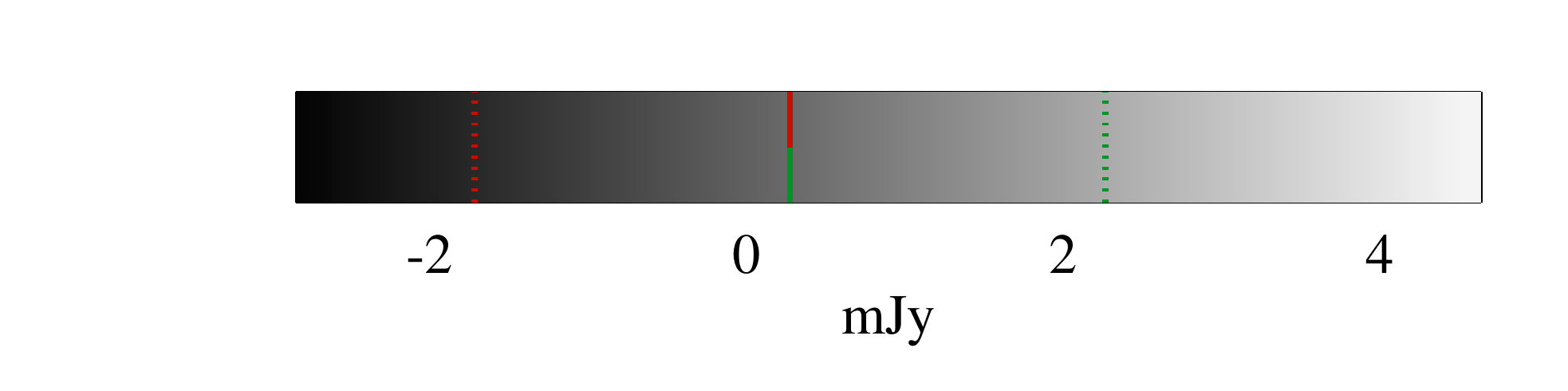}

\caption{The composite images produced by stacking the positions of our IRAC-22.5 sample at $24\,\mu$m (left),  $250\,\mu$m (middle left),  $350\,\mu$m (middle right),  $500\,\mu$m (right). Contours display the same significance levels as in Figure \ref{fig:IRAC_d} except in the $24\,\mu$m  image where we only display a $10 \times$ rms level. The typical source is clearly detected ($>4 \times$ rms) at 24, 250 and 350\,$\mu$m. We carried out this stacking procedure for each of our subsamples, but only present the images for the high stellar mass IRAC-22.5 sample here.}

\label{fig:other stack}
\end{center} 
\end{figure*}

By carrying out a similar stacking analysis at other wavelengths,  we can investigate the spectral energy distribution (SED) of the average high stellar mass source. Firstly, we produce a composite SED of all sources in our IRAC-22.5 sample at optical - NIR wavelengths using the mean magnitudes of our sample in the  \cite{Cardamone10} MUSYC catalogue. Secondly, we use the deep multi-wavelength coverage of the ECDF-S in the MIR and FIR to produce composite images at a number wavelengths covering the dust emission curve at $z\sim3$. We have used the Far-Infrared Deep Extragalactic Legacy Survey  \citep[FIDELS,][]{Dickinson07} deep $Spitzer$ $24\mu$m maps of the field and stack the positions of our sample - obtaining a $\sim12\sigma$ detection. We then apply a similar procedure to the publicly available $Herschel$ SPIRE maps of the ECDF-S at 250, 350 and 500\,$\mu$m taken from the $Herschel$ Multi-tiered Extragalactic Survey \citep[HerMES,][]{Oliver12}. We remove any contribution from background flux in the Herschel maps by performing a Monte Carlo stacking analysis on random source positions in the field (10,000 realisations) and subtracting the median flux over all realisations. After subtracting this background we obtain a $\sim4\sigma$ detection at 250\,$\mu$m and 350\,$\mu$m, and a tentative 2-3$\sigma$ detection at 500\,$\mu$m. 

\begin{figure*}
\begin{center}

\includegraphics[scale=0.28]{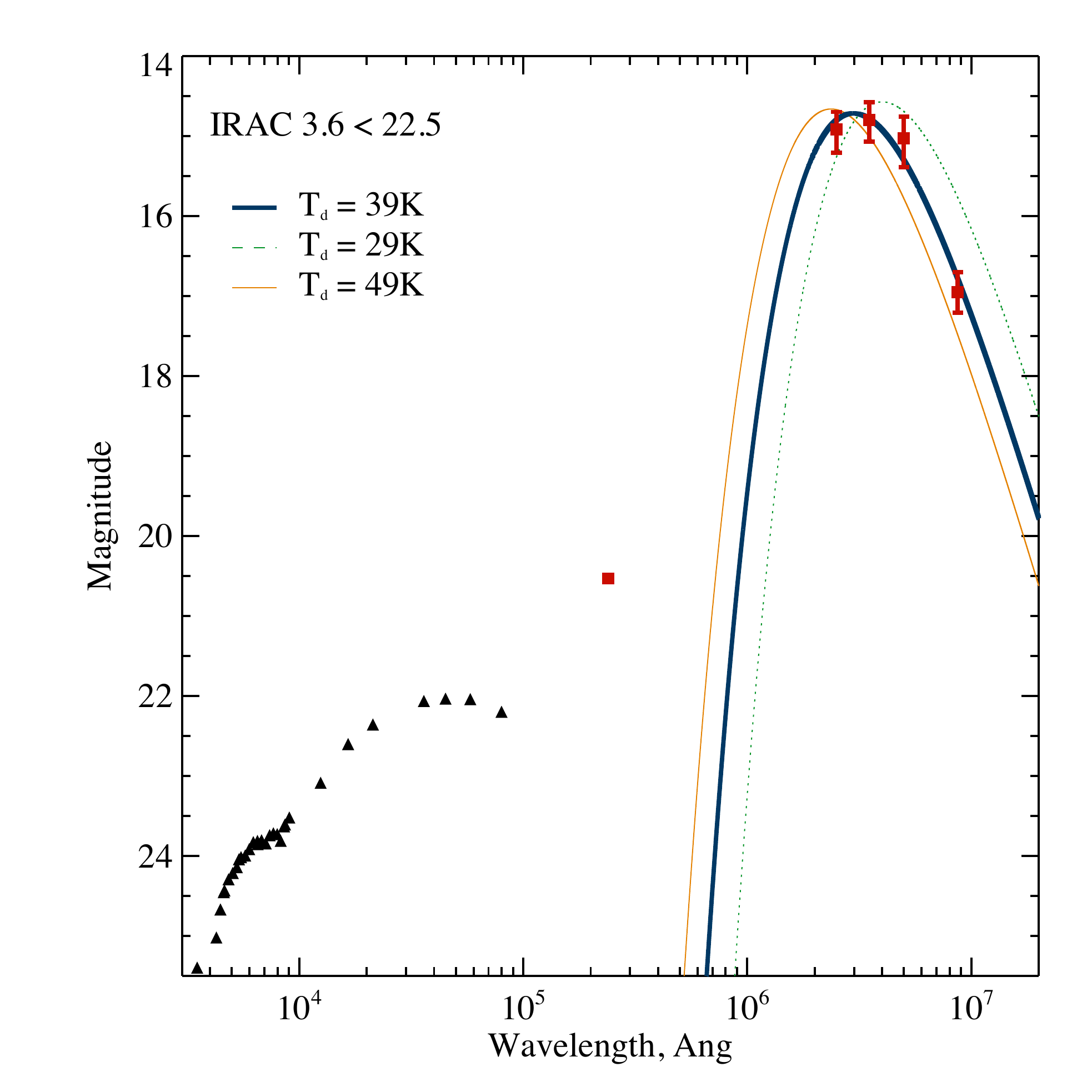}
\includegraphics[scale=0.28]{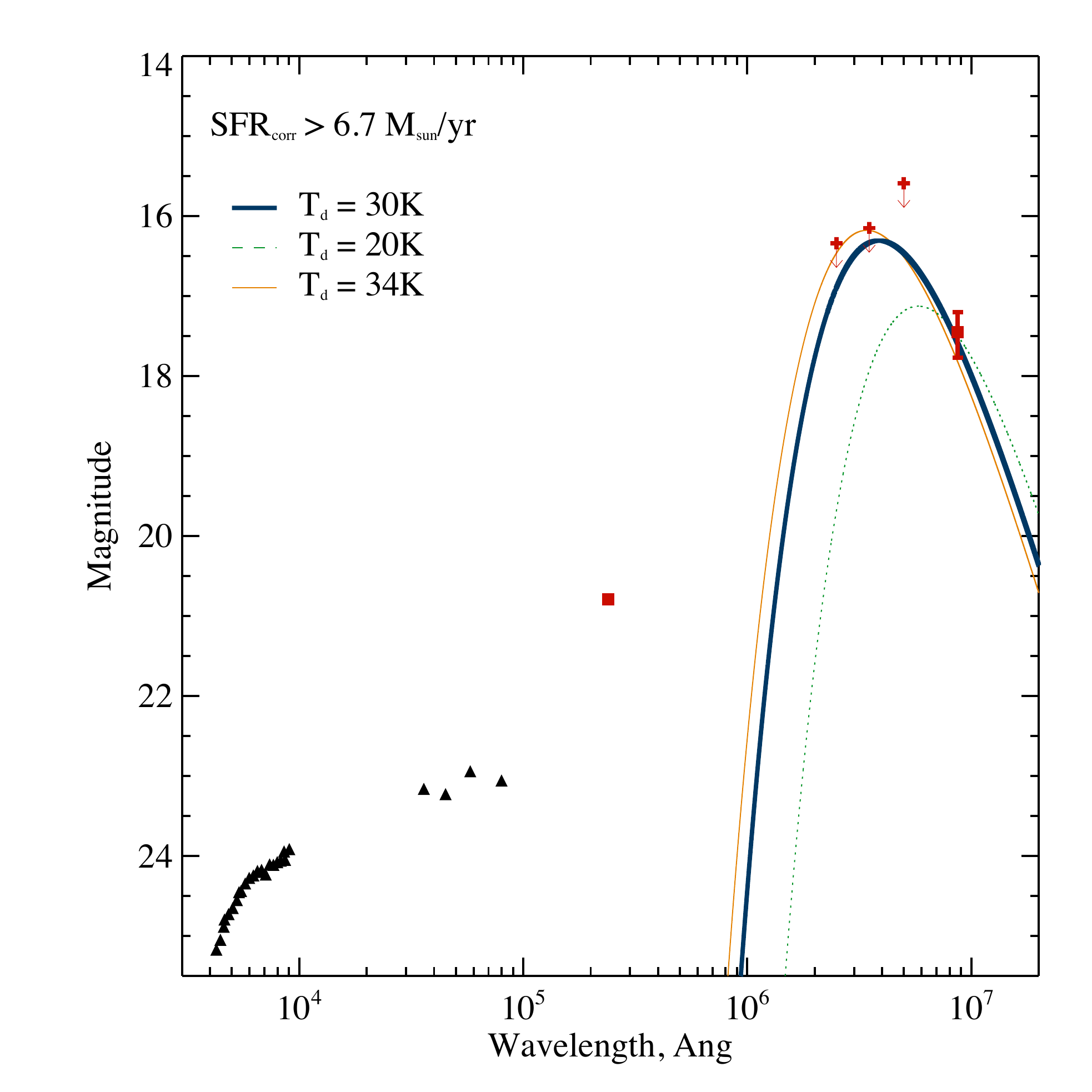}
\includegraphics[scale=0.28]{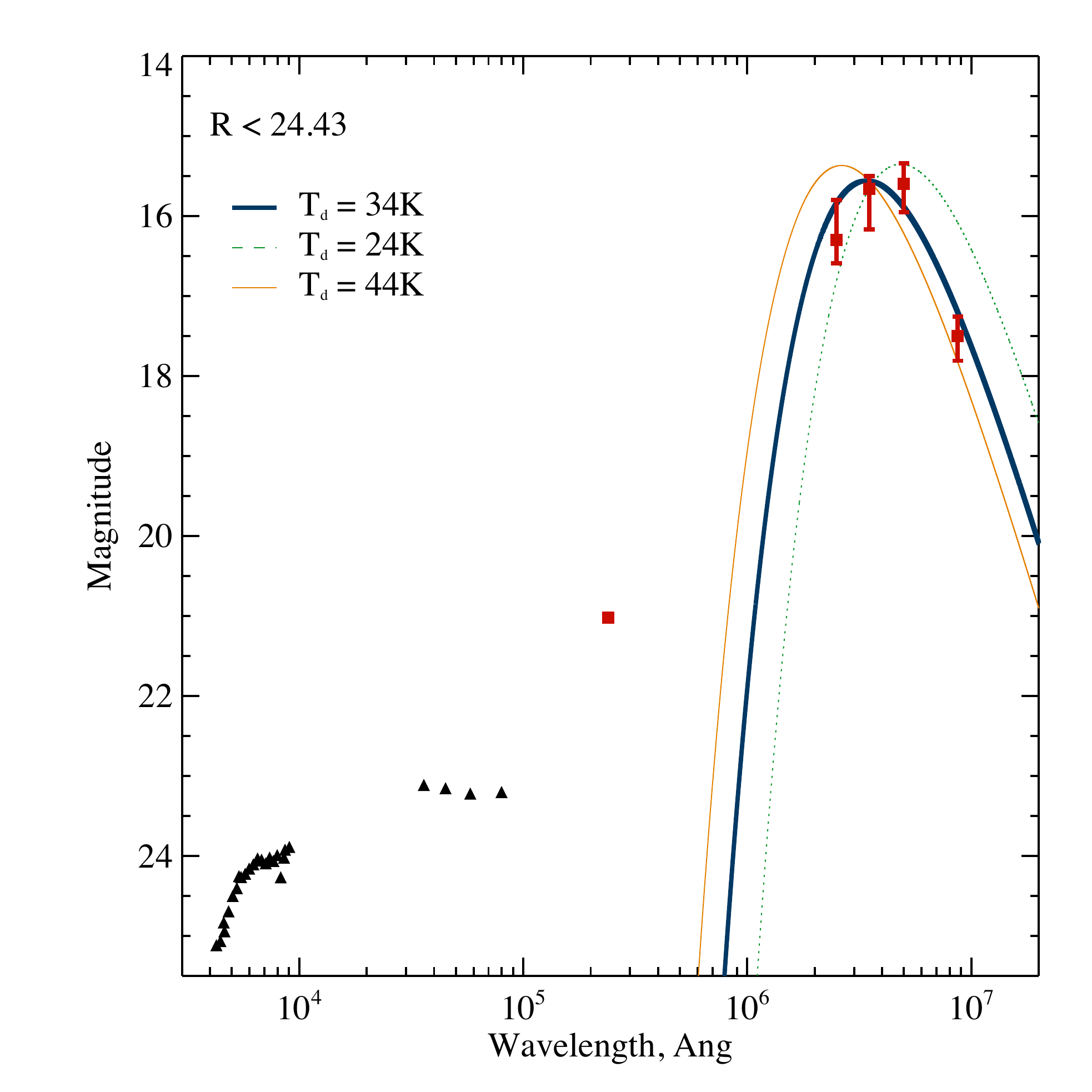}\\

\caption{The composite SED of our IRAC-22.5 sample (left), high-pSFR sample (middle) and UV-bright sample (right). Black triangles display the mean optical and NIR magnitudes taken from the MUSYC catalogue. Red squares and limits display the values obtained from our stacking analysis in this work. The best fit grey body emission curve is plotted as the blue line. For comparison, in the left and right panels, we over plot SEDs with a fixed best-fit $\pm$10K temperature (orange and green lines) - both are found to be inconsistent with the rest-frame FIR data. In the middle panel we plot the SED with an upper limit to the dust teperature which is still consistent with the data (34K). Properties of these SEDs can be found in Table \ref{tab:FIR_props}.  }

\label{fig:SED_IRAC_bright}
\end{center} 
\end{figure*}

Figure \ref{fig:other stack} displays the composite images produced at each wavelength, while Figure \ref{fig:SED_IRAC_bright} (left) displays the composite SED of the composite IRAC-22.5 source in our sample. We fit a grey body emission curve to our FIR data points, assuming a dust SED with power law emissivity, $\beta_{FIR}$\,=\,2, between 20-60K and find a best fit temperature of 39$^{+2}_{-3}$\,K. In Figure \ref{fig:SED_IRAC_bright} we over plot SEDs with fixed temperatures at $\pm10$K of the best fit temperature, displaying that significantly higher or lower temperatures are inconsistent with our data. This represents the first realistic constraint on the dust temperature of non-lensed $z\sim3$ UV-selected star-forming galaxies. 

Following the procedure outlined in \cite{Davies12} and references therein, we integrate this grey body over the 8-1000$\mu$m range and infer a typical FIR luminosity of L$_{FIR}\sim10^{12.0}$L$_{\odot}$ and dust mass of M$_{dust}\sim10^{7.7}$M$_{\odot}$ (all FIR properties of our $z\sim3$ samples are given in Table \ref{tab:FIR_props}).

This luminosity is slightly lower than that of all individually detected submm galaxies at $z\gtrsim3$,  which display L$_{\mathrm{FIR}} >10^{12}$L$_{\odot}$ \citep{Chapman09, Negrello10, Conley11}, but is consistent with $z\sim1-2$ SMGs of \cite{Banerji11}, who find L$_{\mathrm{FIR}}\sim10^{11.0-12.5}$L$_{\odot}$. However,  the L$_{FIR}\sim10^{12}$L$_{\odot}$ sources in \cite{Banerji11} show significantly lower dust temperatures than those found here (25-30\,K) and as such they have larger 870$\mu$m fluxes. While this indicates that the typical SED of our high stellar mass objects  is not directly consistent with that of SMGs over a range of epochs, it suggests that massive $z\sim3$ UV-selected systems may represent sources somewhere between submm bright galaxies (with much larger masses) and sub-mm faint typical star-forming galaxies at $z\sim3$ - displaying significant FIR emission, but being easily detectable in the UV (see Section \ref{sec:mass} for further discussion). Using the typical IRAC-inferred stellar masses of our IRAC-22.5 sample (M$_{\star}\sim10^{10.7}$M$_{\odot}$) we obtain M$_{dust}$/ M$_{dust}$+M$_{\star}\sim$0.0009, suggesting that only a small fraction of their baryonic content is in the form of dust.

\begin{table*}
\begin{scriptsize}
\begin{tabular}{c c c c c c c c c c }
\hline
\hline
Sample & S$_{870\mu m}$ & $\beta_{FIR}$ & T$_{dust}$ & Obs L$_{FIR}$ & M$_{dust}$$^{1}$ & Obs SFR$_{FIR}$$^{1}$ & \uline{Predicted}$^{2}$ & Obscured SFR & Predicted S$_{870\mu m}$$^{3}$\\
 & (mJy/beam) & & (K) & (Log[L$_{\odot}$]) & (Log[M$_{\odot}$]) & (M$_{\odot}$\,yr$^{-1}$) & Observed & (\%) & (mJy/beam)\\ 
\hline
IRAC-22.5 &  0.62 $\pm$ 0.14 & 2.0& 39$^{+2}_{-3}$ & 12.0 & 7.7$^{+0.2}_{-0.1}$ & 168$^{+84}_{-73}$ & $1.1^{+0.9}_{-0.4}$ & 83$^{+5}_{-9}$ & 0.68$^{+0.24}_{-0.13}$  \\

high-pSFR & 0.35 $\pm$ 0.08 & 2.0& $<34$ &  11.5 & $>7.5$ & $<56$ & $>2.8$  & $<68$&  $>9.8$ \\

UV bright &  0.37 $\pm$ 0.09 &2.0& 34$^{+1}_{-1}$ & 11.5 & 7.6$^{+0.1}_{-0.2}$ & 60$^{+24}_{-19}$ &  $0.9^{+0.4}_{-0.2}$ & 71$^{+6}_{-9}$ & 0.3$^{+0.04}_{-0.03}$  \\

All $z\sim3$ &  $<$0.09 & 2.0 & 35$^{4}$ & $<10.9$ & $<7.0$ & $<16$ & $>2.6$ & $<68$ & 0.27 \\

\hline
\end{tabular}

\end{scriptsize}
\caption{The FIR properties of our $z\sim3$ samples. $^{1}$ Properties derived from the best fit SED dust temperature. $^{2}$ The ratio of observed to UV predicted FIR SFRs.$^{3}$ Predicted 870$\mu$m flux, assuming the UV predicted obscured SFR given in Table \ref{tab:UV_props} and best fit SED dust temperature. $^{4}$ As we have no detections with which to constrain the dust temperature of our full sample, it is assumed to be 35\,K. }
\label{tab:FIR_props}    
\end{table*}

\subsection{The high-pSFR sample}

We apply a similar stacking procedure using the LESS data to our high-pSFR sample of the 200  spectroscopically confirmed sources with the highest extinction corrected SFRs. We obtain 0.35\,mJy (formally $\sim4.3\times$ rms) detection near the central position of the stack (bottom left panel of Figure \ref{fig:IRAC_d}), although there is a small (just less than a beam-width) offset between the centroid of the detection and the central position of the stack. We note that there is overlap between the sources in the IRAC-22.5 and high-pSFR samples. However, If we remove these, we still obtain a detection at a slightly lower signal to noise, but consistent with Poisson statistics. The measured flux density in our high-pSFR sample is somewhat lower than that obtained for the IRAC-22.5 sample ($\sim50\%$), but is detected at the similar significance due to the larger sample size. For comparison we note that the typical stellar mass of high-pSFR sample (as measured from the IRAC fluxes) is M$_{\star}\sim10^{10.4}$M$_{\odot}$. Therefore, while these sources have lower stellar masses than our IRAC-22.5 sample, they still display above average mass in comparison to the general $z\sim3$ LBG population \citep{Shapley01}. In conjunction with the IRAC-22.5 sample this highlights the well documented correlation between SFR and stellar mass in high redshift star-forming galaxies \cite[$e.g.$][]{Rodighiero10, Magdis_c} - see section \ref{sec:main} for further discussion.

We consider these sources in both the $Herschel$ SPIRE maps and $Spitzer$ MIPS maps of the field, as above, and find that this sample is undetected in any of the $Herschel$ bands, ruling out temperatures of $>34$K (Figure \ref{fig:SED_IRAC_bright}). However, without detections at multiple wavelengths we can say little else about the true dust temperature. We infer a FIR luminosity of L$_{FIR}\sim10^{11.5}$L$_{\odot}$ and, at T$_{dust} <$ 34K, a dust mass of  M$_{dust}\gtrsim10^{7.5}$M$_{\odot}$. Using the IRAC-inferred stellar masses we obtain M$_{dust}$/ M$_{dust}$+M$_{\star}\gtrsim$ 0.0008 - consistent with the IRAC-bright sample.

\subsection{The full and UV-bright sample}

These two samples represent a standard LBG selection typical of many previous studies of $z\sim 3$ LBGs, and a higher unobscured UV-luminosity cut of the same sample. The latter contains a mix  of sources with either more ongoing star formation, or less extinction towards their star forming regions than the former.We apply our 870$\mu$m stacking procedure to both samples and find no detection for the full sample (to a limit of $<0.09$ mJy). A similar stacking of the $Herschel$ SPIRE data for these source again results in non-detections. The properties of this sample are summarised in Tables \ref{tab:UV_props} and \ref{tab:FIR_props}, assuming, in the absence of any detection in the FIR a dust temperature of 35K, and the image of the 870$\mu$m stack is shown along with those of the full samples at higher redshift in Figure \ref{fig:phot_stacks}.

Stacking the 870$\mu$m data for the  UV-bright sample of the 200 R-band brightest $z\sim$3 LBGs we obtain an average flux of 0.37\,mJy ($\sim4\times$rms, bottom right panel of Figure \ref{fig:IRAC_d}). We note again that there is overlap between our IRAC-22.5 and UV-bright samples (20 sources). If we remove these sources we still obtain a detection once again at a slightly lower signal to noise, but still consistent with Poisson statistics. The measured flux density in our UV-bright sample is almost identical to that obtained for the high-pSFR sample. The typical stellar mass of UV-bright sample is M$_{\star}\sim10^{10.35}$M$_{\odot}$ - also similar to the high-pSFR sample. In fact, the only significant difference between our high-pSFR and UV-bright samples are their UV-spectral slopes ($\beta_{UV}=-1.4$ and $\beta_{UV}=-1.8$ respectively, with mean spectral slope error on an individual source of $\sim \pm0.06$). The almost identical 870$\mu$m flux is surprising given that the UV-spectral slope is thought to be indicative of extinction, with the extincted flux re-emitted in the FIR \citep[$e.g.$][]{Meurer99, Adelberger00, Finkelstein09}. The lack of variation of observed 870$\mu m$ flux with these samples, potentially suggests a lack of correlation between UV-spectral slope and FIR emission (at least for sources with the largest deviation from $\beta_{UV}=-2$) and that the actual total star formation rate in our supposed high-pSFR sample is no higher than for our UV-bright sample.  This will be discussed further in Section \ref{sec:uv_slope}.

We once again consider these sources in both the $Herschel$ SPIRE maps and $Spitzer$ MIPS maps of the field. Figure \ref{fig:SED_IRAC_bright} (right) displays the composite SED of our UV-bright sample. We fit a grey body emission curve to our FIR data points, assuming a dust SED with power law emissivity $\beta_{FIR}$\,=\,2, and find a best fit temperature of 34$\pm1$\,K. We note that this temperature is consistent with that used in our previous study of $z\sim5$ sources \citep{Davies12}, strengthening the validity of that analysis. We over plot SEDs with fixed temperature at 24\,K and 44\,K, once again displaying that significantly higher or lower temperatures are inconsistent with our data.  At 34\,K we infer a FIR luminosity of L$_{FIR}\sim10^{11.5}$L$_{\odot}$ and dust mass of  M$_{dust}\sim10^{7.6}$M$_{\odot}$ for the composite source in our UV-bright sample. Using the IRAC-inferred stellar masses we obtain M$_{dust}$/ M$_{dust}$+M$_{\star}\sim$0.0017 - consistent with the high-pSFR sample and the high stellar mass IRAC-22.5 sample.

\section{Results for higher redshift LBGs}

At higher redshifts LBGs are likely to display much lower FIR fluxes, assuming the same SED shape as our detected $z\sim3$ samples - while the increasing luminosity distance and inverse K-correction provide roughly equal and opposite scaling to the observed flux at higher redshifts,  LBGs at $z\gtrsim3$ are significantly less massive than those at $z\sim3$ \citep{Verma07}. Therefore, are likely to be less FIR luminous assuming the potential $L_{FIR}$-stellar mass correlation discussed previously. Coupled with the smaller sample sizes, it is unlikely that the typical $z>3$ LBG will be detected in our composite images. 

\begin{figure*}
\begin{center}

\includegraphics[scale=0.3]{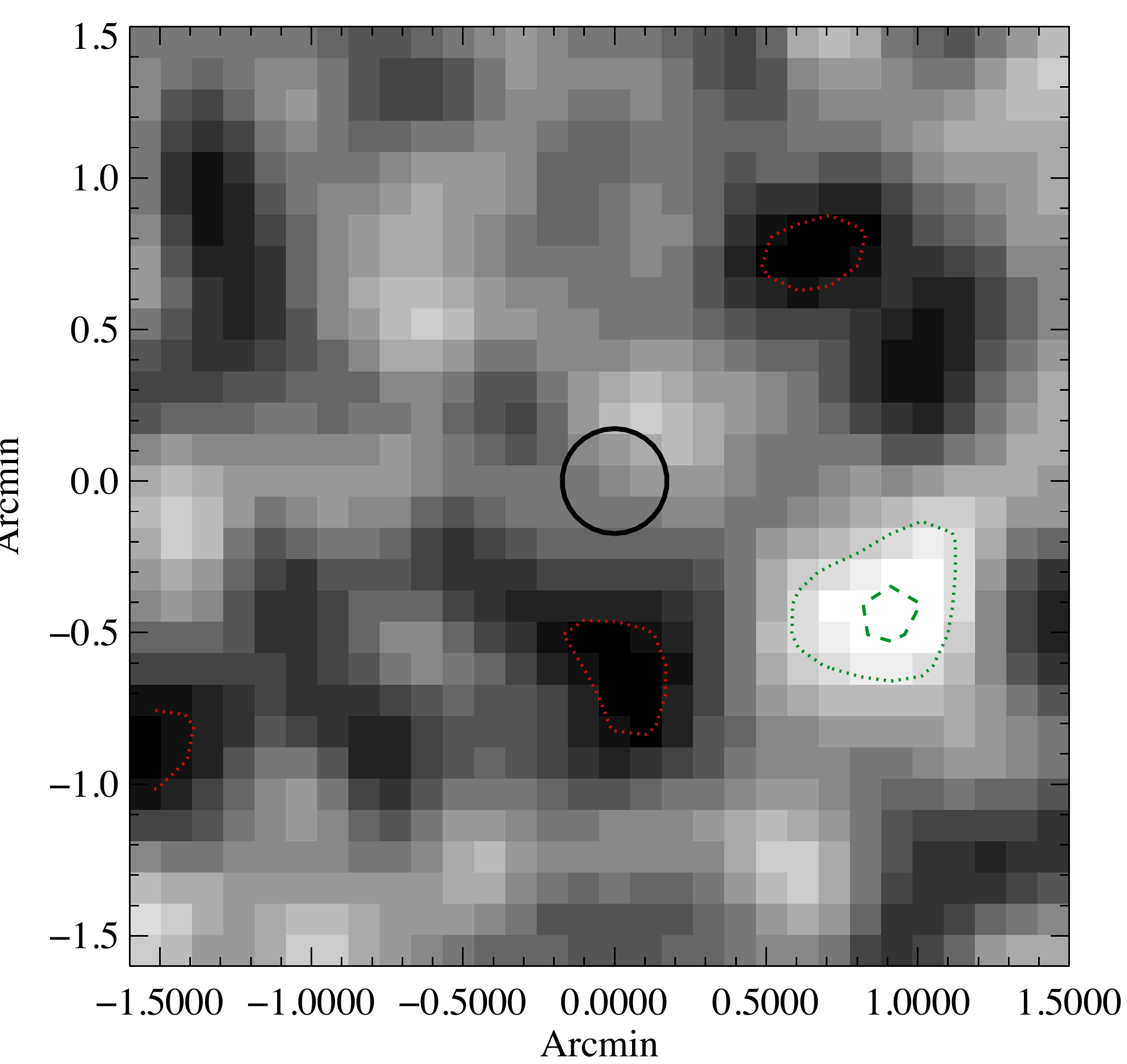}
\includegraphics[scale=0.3]{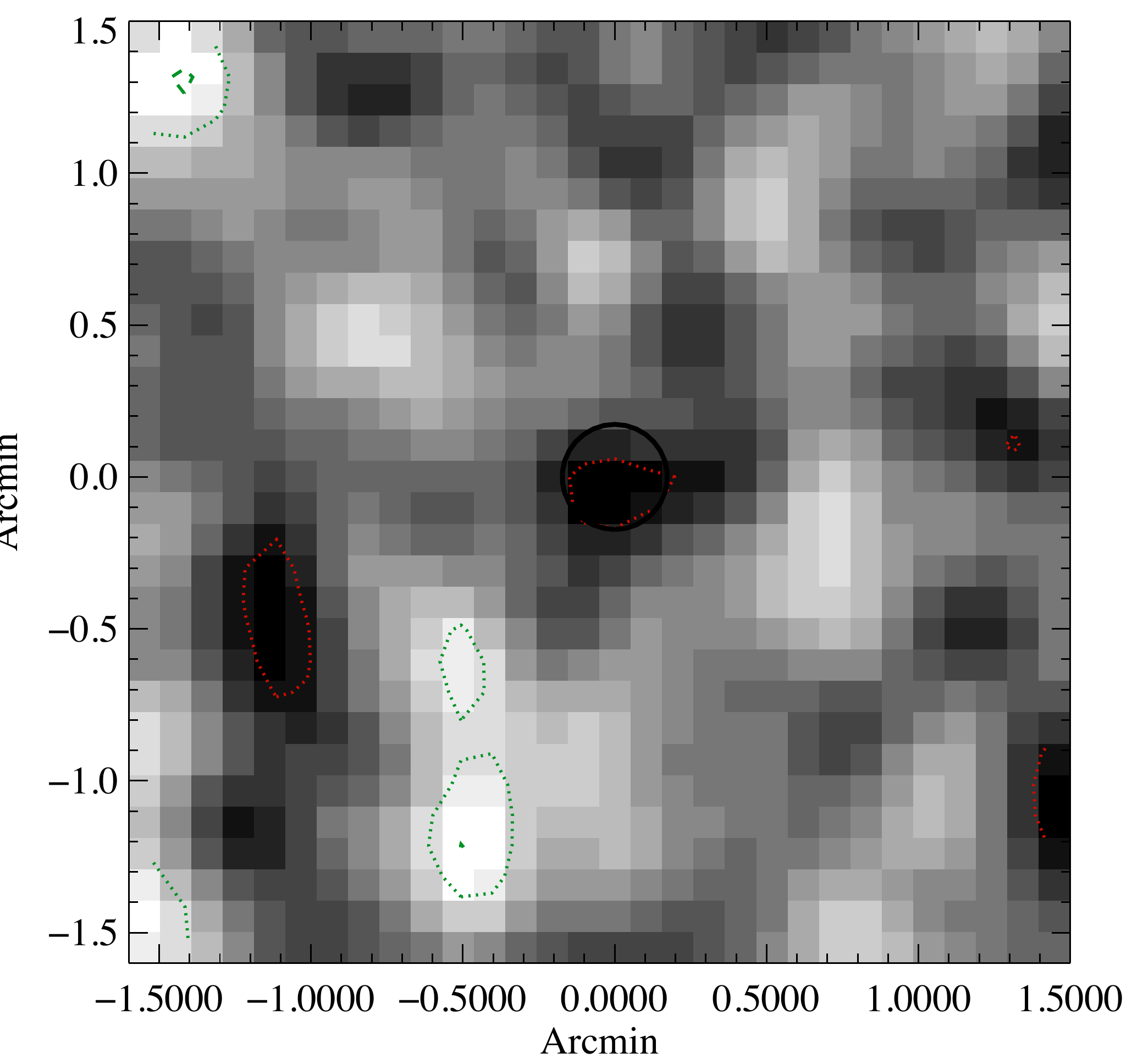}\\

\includegraphics[scale=0.3]{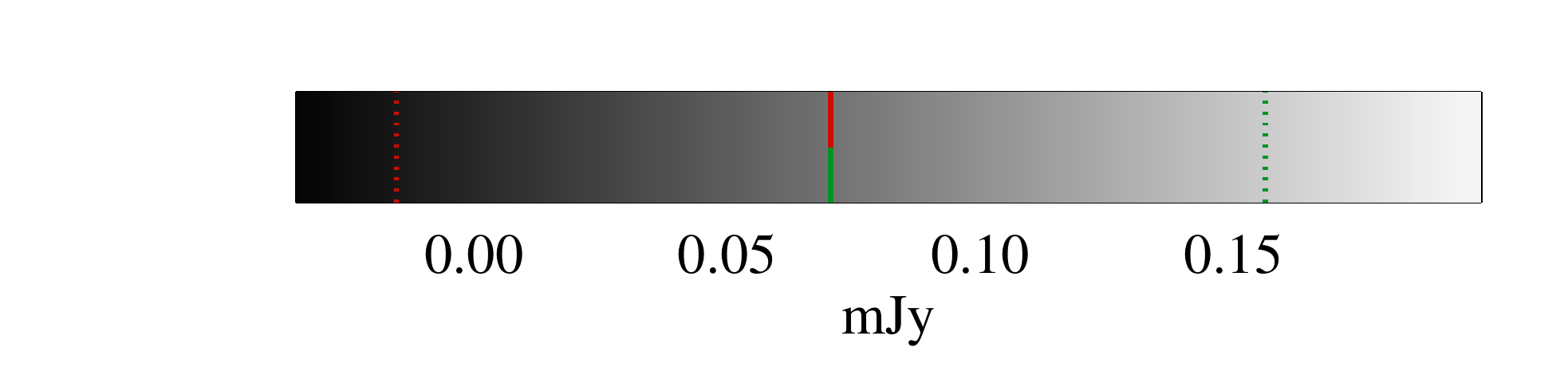}
\includegraphics[scale=0.3]{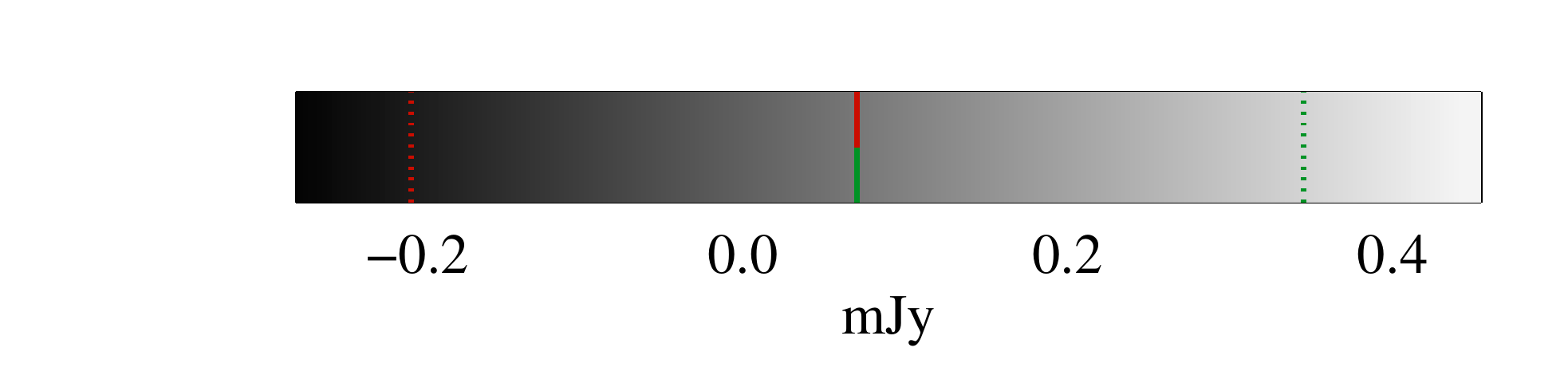}\\

\includegraphics[scale=0.3]{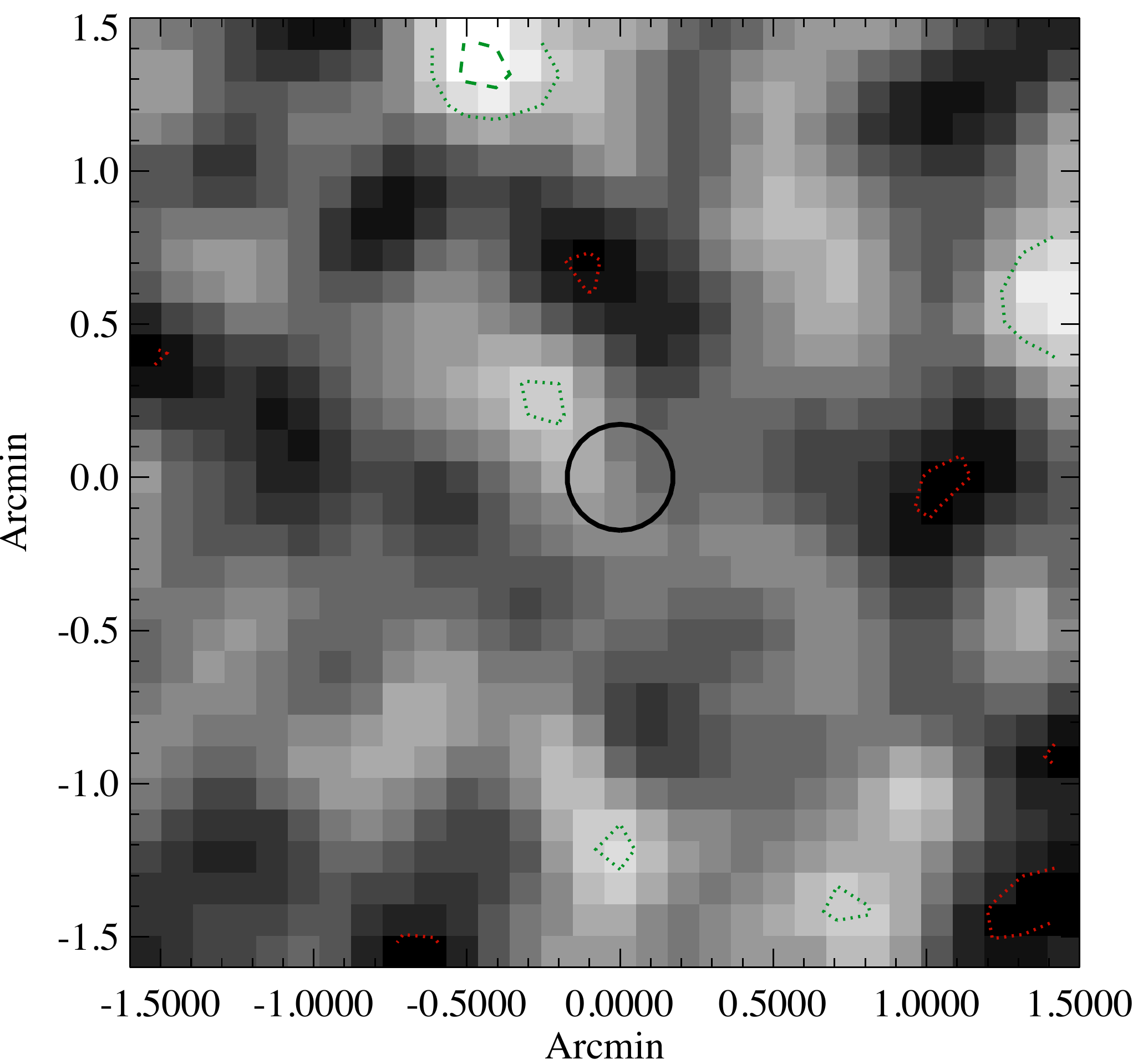}
\includegraphics[scale=0.3]{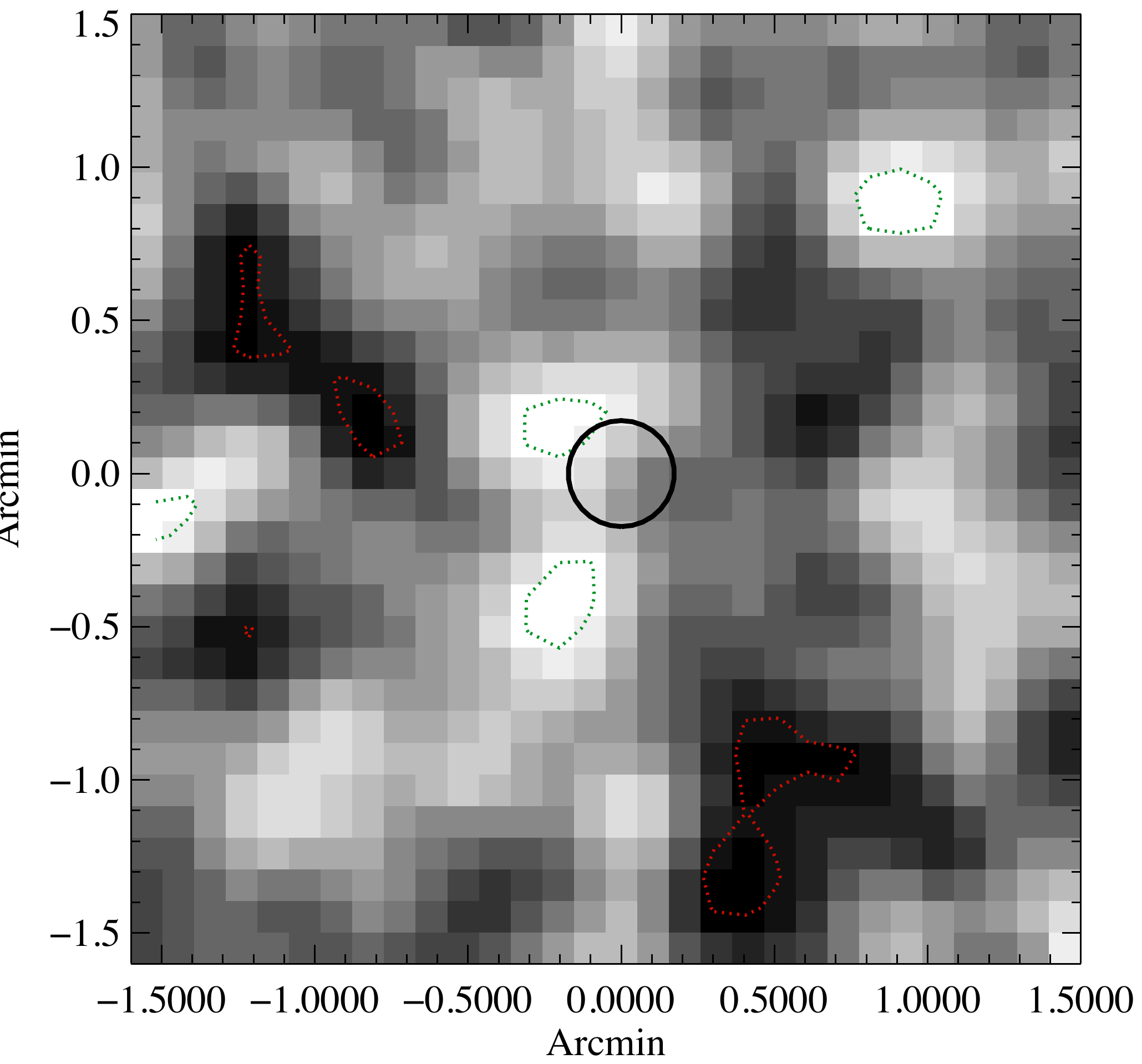}\\

\includegraphics[scale=0.3]{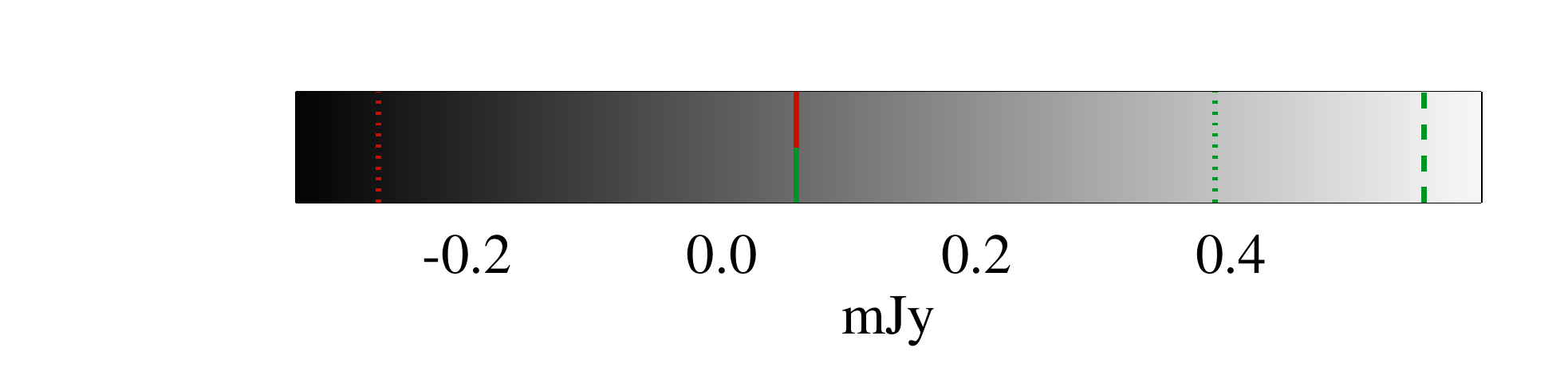}
\includegraphics[scale=0.3]{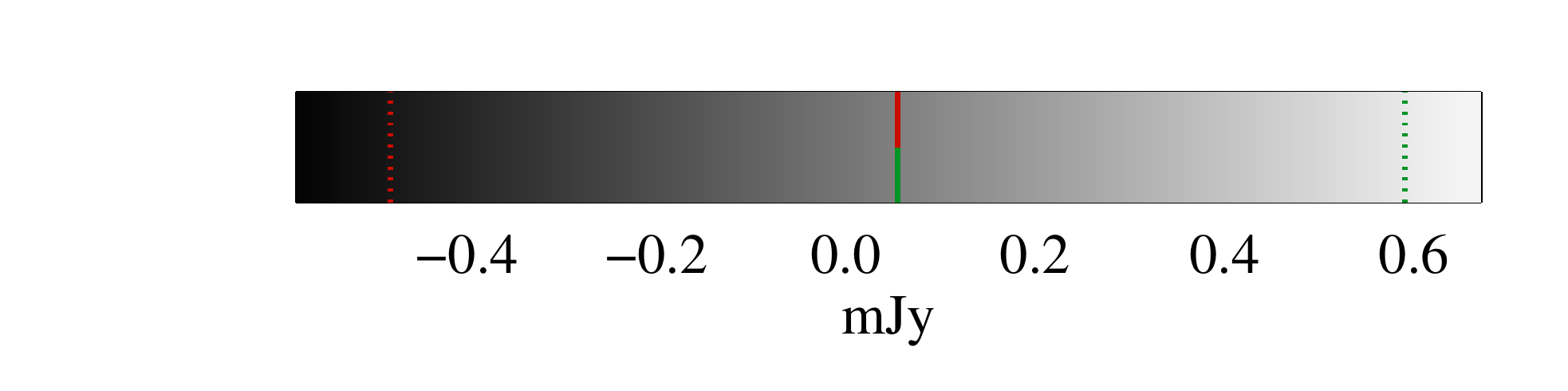}

\caption{Composite 870$\mu$m images of the positions of our $full$ samples of high redshift star-forming galaxies in the ECDF-S at $z\sim3$ (top left), 4 (top right), 4.5 (bottom left) and 5 (bottom right) convolved with a gaussian of the same FWHM as the LABOCA beam size. Contours show both positive (green) and negative (red) deviations of 2 (dotted line) and 3 (dashed line) $\times$ rms away from the mean value in the field. The LABOCA beam size is displayed as the black circle.}

\label{fig:phot_stacks}
\end{center} 
\end{figure*}

Figure \ref{fig:phot_stacks} displays the stacked images produced from our complete LBG samples at $z\sim 3$ (discussed above), $z\sim4$ and 5, and the LAE sample at $z\sim4.5$. We find that no source is detected at a $\gtrsim2\times$ rms level in the centre of any of the composite images. We do obtain a $\sim2\sigma$ detection close to the central region in our $z\sim5$ sample. However, this is consistent with the noise characteristics of the data. Table \ref{tab:props}  displays the properties of the composite source at each redshift, derived from our stacked images. These non-detections are consistent with our dust SED shape for the UV-bright sample (which is likely to be most appropriate at $z>3$) being applicable to LBGs at $z\,>\,3$, only scaled in UV luminosity and stellar mass.          

\vspace{2mm}

As the typical LBGs remains undetected in our composite images at all epochs, we note that observations much deeper than those previously obtained are required to individually detect these systems (reaching $<$0.2\,mJy). Observations such as these are impractical with single dish observatories and therefore with require observations with high sensitivity interferometers - $e.g.$ the fully operational ALMA.

In combination, our results suggest that high redshift LBGs galaxies have low dust content ($<0.1\%$ of their total baryonic mass). This result, coupled with the lack of large quantities of molecular gas in these systems \citep{Livermore12,Magdis12}, suggests that they do not have large quantities of baryonic material which is not observable in the rest frame UV. This has important consequences for the nature of the LBG phenomenon, suggesting that they are independent star-burst galaxies with little obscured material and not low extinction sight-lines through a much larger obscured system \citep[see ][for discussion]{Davies10,Davies12}.

\begin{table*}
\centering

\begin{scriptsize}

\begin{tabular}{c c c c c c c c c c}
\hline
\hline
Redshift & Sources$^{1}$  & N$_{\mathrm{gals}}$ &  S$_{870\mu m}$ & $\beta_{FIR}$& T$_{dust}$ & L$_{\mathrm{FIR}}^3$ & M$_{\mathrm{dust}}^4$& $\mathrm{\dfrac{M_{Dust}}{M_{Dust}+M_{\star} }}^{5}$& SFR$^{6}$\\
& & &(mJy/beam) & & (K) & (Log[L$_{\odot}])$& (Log[M$_{\odot}$])& & (M$_{\odot}$\,yr$^{-1}$) \\

\hline

3.6 - 4.5 & V-drops & 68&0.35 & 2.0& 35& $<$11.5 &$<$7.1 & $<$0.006&$<$56\\
4.5 & LAEs & 46& 0.40 &  2.0& 35&$<$11.6 & $<$7.1 & $<$0.112&$<$62\\
4.7 - 5.5 & R-drops & 20&  0.61 & 2.0& 35& $<$11.7 & $<$7.3 &  $<$0.012&$<$92\\
 
\hline
\end{tabular}
\end{scriptsize}
\caption{The properties of the typical LBG in our samples at each redshift. $^{1}$ Source types: V-drops and R-drops: LBGs samples at $z\sim4,$ and 5 selected in this study and  LAEs: Lyman-$\alpha$ Emitters at $z\sim4.5$ identified in \citet{Zheng11}. 
$^{2}$ 2$\times$rms limits including the 0.072\,mJy\,beam$^{-1}$ scaling factor used to account for source confusion.
$^{3}$ FIR Luminosity limit derived for our $2 \times$ rms (+ 0.072\,mJy\,beam$^{-1}$) limit.
$^{4}$ Dust mass at a $2 \times$ rms (+ 0.072\,mJy\,beam$^{-1}$) limit.
$^{5}$ The cool dust to stellar mass fraction. Stellar masses are taken as the mean mass of a typical source at  $z\sim4$ ($\sim10^{9.3}$M$_{\odot}$, Hathi et al, in prep), $z\sim4.5$  \citep[LAEs\,$\sim10^{8.0}$M$_{\odot}$,][]{Finkelstein07} and $z\sim5$ \citep[$\sim10^{9.2}$M$_{\odot}$,][]{Verma07}. 
$^6$ The obscured star formation rate derived using the the Kennicutt relation \citep{Kennicutt98}.}

\label{tab:props}        
\end{table*}

\section{Discussion}

\subsection{Comparisons to other populations}
\label{sec:compare}

As the detections at $z\sim 3$ represent some of the first multi-frequency constraints to the dust SED of the population of  unlensed LBGs, it is interesting to consider the FIR properties of these sources in comparison to other populations of galaxies. The most direct comparison can be made with the recent detection of FIR emission from LBGs at $z\sim4$. As noted previously, \cite{Lee12} stacked the $Herschel$ SPIRE data at the positions of $z\sim4$ LBGs in the Bootes field. They obtain a detection of the rest-frame UV-brightest sources at 350 and 500$\mu$m. Their detected sample consists of sources with $I_{AB}\,<\,$24.3, hence is roughly equivalent to the $R$-band magnitude limits of our UV-bright sample (both represent an observation of the rest-frame UV-continuum at their respective redshifts). Their FIR fluxes are 1-2\,mJy and have best-fit SEDs with dust emission peaking at $\gtrsim100\,\mu$m in the rest frame of the galaxy. Our results on the  $z\sim3$ LBGs are consistent with these, assuming that there is no  evolution in the physical properties of galaxies selected to have similar rest-UV properties, with the dust SED of our UV-bright sample reaching a maximum of $2.1\pm0.58$\,mJy and potentially rising in emission out to 500\,$\mu$m.

 \begin{figure*}
\begin{center}

\includegraphics[scale=0.6]{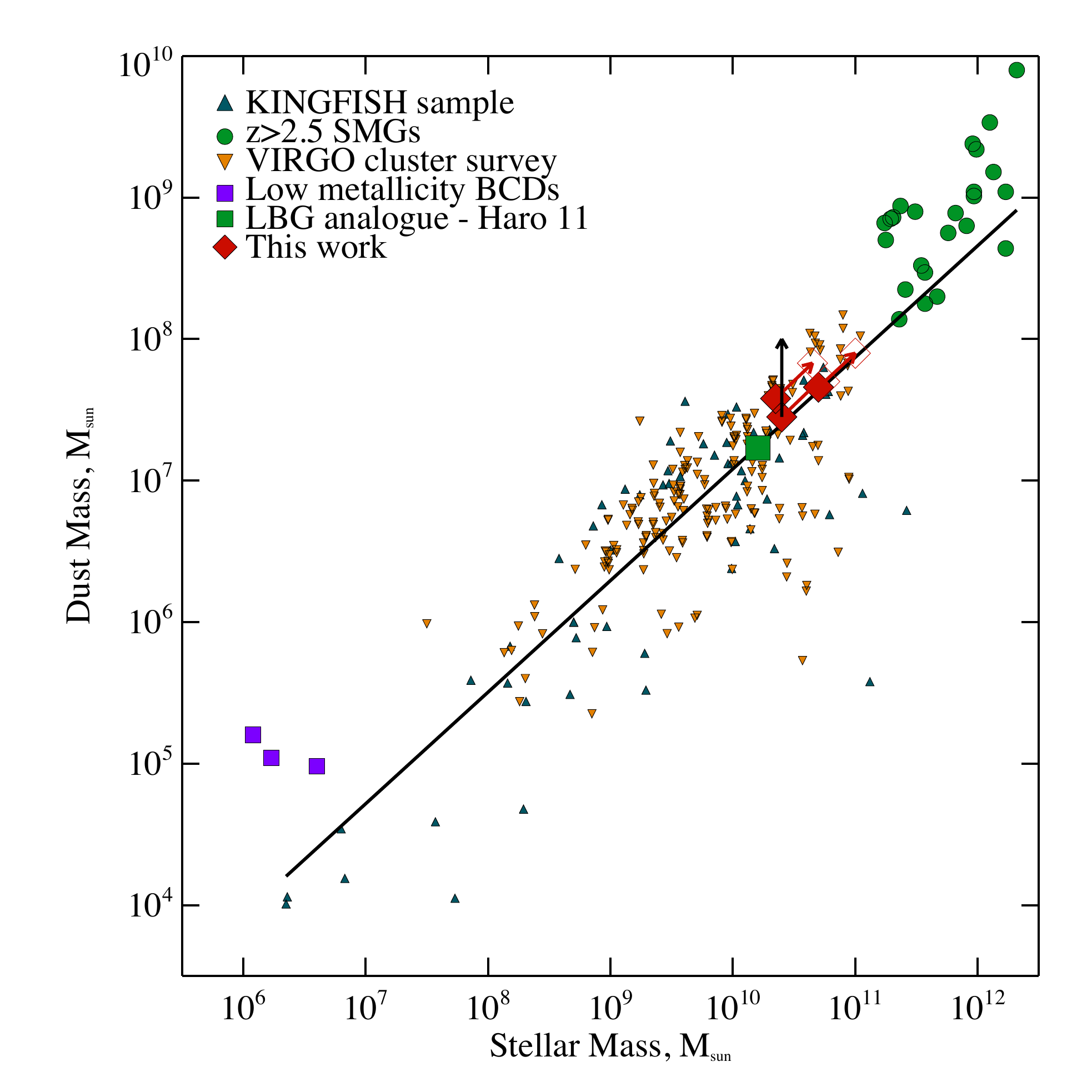}\\

\caption{The stellar mass to dust mass relation. Our IRAC-22.5, high-pSFR (black arrow, lower limit to dust mass as temperature is not constrained) and UV-bright  samples are displayed as the filled red diamonds. Sources from the $Herschel$ KINGFISH survey \citep{Skibba11} are displayed as blue triangles, $Herschel$ VIRGO cluster survey \citep{DaviesJ12, Auld12} as orange triangles, $z>2.5$ submm sources \citep{Michalowski10} as green circles and low metallically blue compact dwarfs \citep[BCDs][]{Hunt05} as purple squares. The green square displays the well studies local LBG analogue - Haro 11 \citep[$e.g.$][]{Heckman05, Galametz09}. The black line displays the best fit linear correlation to the combined data sets. Red arrows/open red diamonds display the limit to the  growth of the composite system in our subsamples, assuming roughly half of their baryonic mass is in the form of molecular gas, a 100$\%$ efficiency in converting molecular gas into stars and that the sources would evolve following the same M$_{dust}$/M$_{*}$ ratio as the best fit linear correlation.}

\label{fig:mass}
\end{center} 
\end{figure*} 
 
It is also interesting to consider whether or not these systems display similar dust/stellar mass characteristics to those selected for their high submm luminosity,  and whether they could potentially evolve into such systems at a later epoch. Figure \ref{fig:mass} shows  the stellar mass against dust mass for our detected subsamples. We also show the properties of low redshift sources from the $Herschel$ KINGFISH survey \citep{Skibba11} and $Herschel$ Virgo cluster survey \citep{DaviesJ12, Auld12}, $z>2.5$ submm sources \citep{Michalowski10} and metal poor blue compact dwarf galaxies \citep[BCDs,][]{Hunt05}. Our samples fall very close to the linear best fit to the data and are consistent with, but at the high mass end, of the low redshift systems. While they fall far below the SMGs both in dust and stellar mass, they do display similar dust fractions. Interestingly, our LBGs display significantly lower dust to stellar mass fractions than those found for the local BCDs. Although significantly less massive, these BCDs have specific star-formation rates and metallicities which are similar to LBGs and have been proposed as scaled down LBG analogues at low redshift. The difference in M$_{dust}$/M$_{*}$ ratio between these sources and our LBG samples suggests that care must be taken in comparing BCDs with LBGs, as they may display distinctly different dust characteristics. We note that the most well studied low redshift LBG analogue \cite[Haro II, see][, etc]{Heckman05, Galametz09} displays an almost identical stellar to dust mass ratio to our subsamples - suggesting that Haro 11 may have similar dust properties to $z\sim3$ LBGs. However, \cite{Galametz09} find a dust SED for Haro 11 peaking at $\sim40\,\mu$m in the rest-frame. This corresponds to a dust SED peaking at $160\,\mu$m for a identical source as $z\sim3$ and is inconsistent with our LBG samples - it would require a dust SED which peaks below our $Herschel$ 250\,$\mu$m point, which is unlikely given the rising flux in both our SEDs between 250 and 350\,$\mu$m. This suggests that Haro 11 displays a much higher dust temperature than our $z\sim3$ LBGs. This is intriguing given that Haro 11 shares many other characteristics with LBGs at $z\sim3$ and may potentially display differences between the ISM topography in Haro 11 and our high redshift sources.

Recently a number of studies have investigated the molecular gas content of LBGs  \cite[$e.g.$][]{Livermore12, Magdis12} finding that, consistently over a number of sources, the molecular gas content of LBGs is roughly the same as their stellar mass ($i.e.$ the stellar and molecular gas content each make up $\sim50\%$ of the total baryonic mass). Assuming a similar molecular gas fraction in our systems, and the same stellar to dust mass scaling, we predict their potential growth through star-formation (irrespective of merging). The red arrows in Figure \ref{fig:mass} display the maximum growth of the systems assuming 100$\%$ efficiency in the conversion from molecular gas to stars, no accretion of cool gas onto the galactic halo from filaments in the the large scale structure, and evolution with the same M$_{dust}$/M$_{*}$ ratio ($i.e$ following the same slope as the observed linear correlation for low redshift sources). We find that at best these systems will reach $\sim10^{11}$M$_{\odot}$, and exhaust all of their molecular gas in $\lesssim500$Myr. Therefore, despite displaying moderate FIR fluxes,  it is unfeasible that these systems will grow into submm bright galaxies through the conversion of all of the available material into stars - they would require a significant number of major mergers to reach the required mass. However, this does not rule out our samples (specifically the IRAC-22.5 sample) representing the very low mass end of the submm galaxy distribution, which falls below the individual source detection limit of submm observatories (see Section \ref{sec:mass}).     

Lastly in this section, we consider the dust temperature of our composite sources - 39$^{+2}_{-3}$\,K for the high stellar mass IRAC-22.5 sample, $<34$\,K for the high-pSFR sample and 34$^{+1}_{-1}$\,K for the UV-bright sample. \cite{Lee12} estimate dust temperatures of $\sim30$K for $z\sim4$ LBGs, although they only obtain detections short-ward of the dust peak and hence can not accurately constrain the temperature. We note that our dust peak position is largely constrained by our 870\,$\mu$m detection - producing a best fit SED to higher temperatures than the \cite{Lee12} result (for reasonable assumptions of power law emissivity). \cite{Reddy12} also obtain a best fit temperature of $\sim30$K for systems at $z\sim2$, although they also do not have a detection long-ward of the dust peak. More generally, if we once again compare our composite sources with other galaxy populations, we find that our dust temperatures are more consistent with the submm bright sources \citep[mean $\sim41$K,][]{Michalowski10} and ``power-law'' SED type distant obscured galaxies (DOGs) at $z\sim2$ \citep[median $\sim35$K,][]{Melbourne12} than with $z\sim1-2$ SMGs \citep[25-30\,K,][]{Banerji11} and local galaxies - mean $\sim27$K \citep[][although calculated for a $\beta_{FIR}$=1.5 model, they state that assuming $\beta_{FIR}$=2.0 only produces slightly lower temperatures]{Skibba11} and $\sim$20\,K \citep{DaviesJ12, Auld12}.

In summary, we find that our comparatively high stellar mass IRAC-22.5 sample shares many characteristics with submm bright sources at the the same redshift, albeit at much lower masses. They have similar stellar to dust mass ratios and dust temperatures. These systems may therefore represent the low mass end of the submm galaxy distribution (see Section \ref{sec:mass}). Our other subsamples appear to have lower dust temperatures, more consistent with previous estimates for typical star-forming galaxies at $2<z<4$, but higher than $z\sim1-2$ SMGs and local galaxies.

\subsection{Total SFRs and the `main sequence' of star-forming galaxies}
\label{sec:main}

These results provide a direct measurement of these systems' FIR luminosity, and hence obscured SFR. Previous studies have attempted to constrain total SFR (obscured + unobscured) in high redshift galaxies by correcting unobscured SFRs - applying a reddening model to the observed UV-optical emission \cite[$e.g.$][]{Magdis_c}. Here, we provide a more direct measurement of the obscured star-formation, allowing us to constrain the total SFR in our samples of $z\sim3$ systems.

We use our integrated L$_{\mathrm{FIR}}$ to obtain a FIR-derived SFR limit, using the Kennicutt relation for local starburst galaxies \citep{Kennicutt98}:

 \begin{equation}
\mathrm{SFR_{FIR}\, (M_{\odot}/yr)=4.5 \times 10^{-44}\,L_{FIR} \, (ergs/sec)} ,
\end{equation}

We derive a typical obscured SFR for our stellar-mass selected IRAC-22.5 sample $\sim170$M$_{\odot}$/yr (see Table \ref{tab:FIR_props}). If we once again use the conversion between UV flux and SFR (equation \ref{eq:UV_SFR}), we obtain a typical unobscured SFR of $\sim34$\,M$_{\odot}$yr$^{-1}$, hence a total SFR of $\sim200$\,M$_{\odot}$yr$^{-1}$.  Applying the same procedure to the high-pSFR and UV-bright samples we obtain a total SFR of $<80$M$_{\odot}$/yr and $\sim85$M$_{\odot}$/yr (see Tables 2 and 3 for individual breakdown). For at least the UV-bright and the IRAC-22.5 samples the majority (upto $\sim80\%$) of the star-formation is obscured.   As noted earlier, the behaviour of the sample selected to have the highest star formation rate (assuming the slope of the UV SED reflects significant extinction and we can accurately characterise that extinction) is unexpected. For any reasonable dust temperature, it appears to have a lower total SFR to  both the UV-bright and stellar mass selected samples. We discuss this further in Section \ref{sec:uv_slope}.

\begin{figure}
\begin{center}

\includegraphics[scale=0.42]{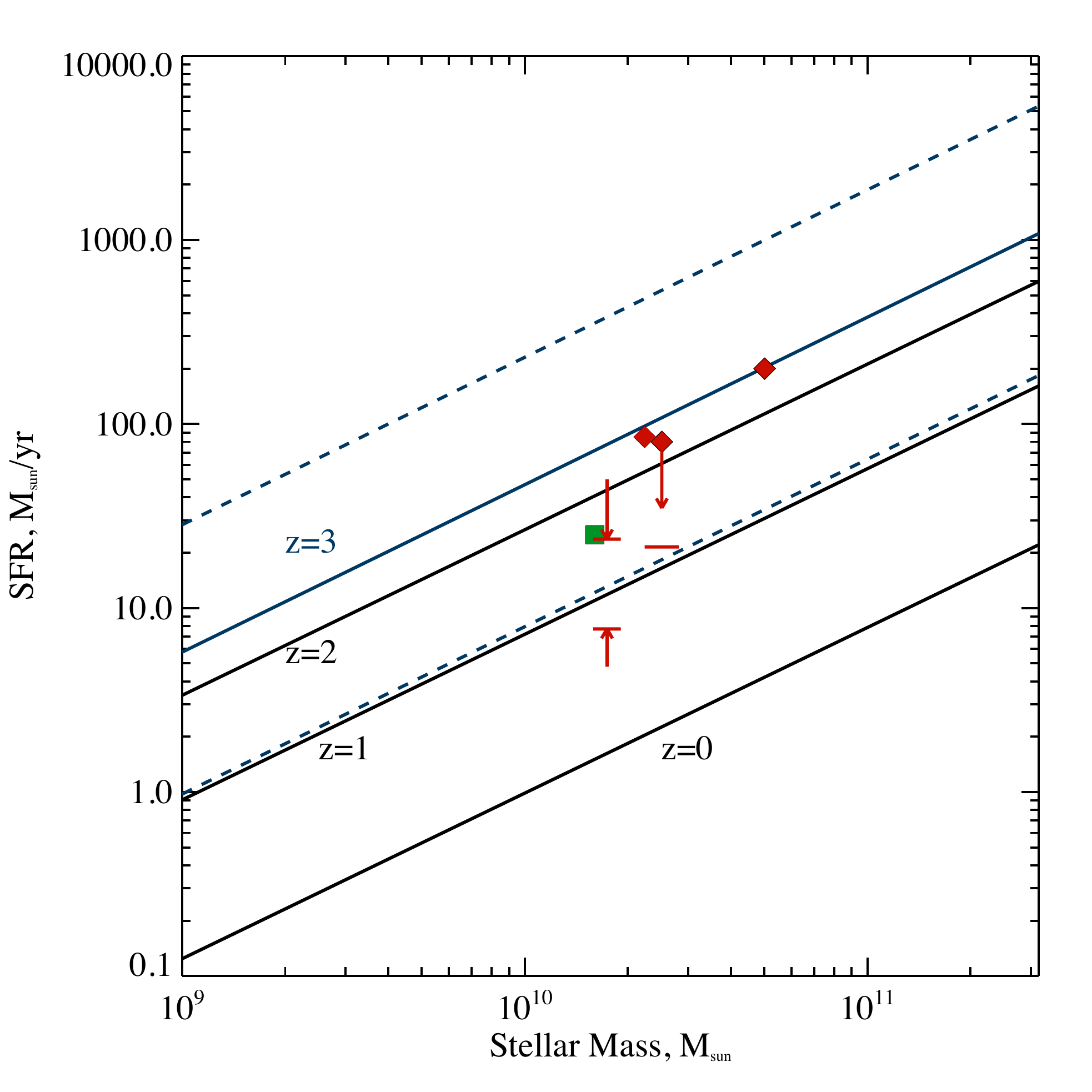}

\caption{The SFR against stellar mass correlation for star-forming galaxies. Red diamonds display the positions of our IRAC-22.5, high-pSFR and UV-bright sample. Stellar masses are calculated from the mean IRAC fluxes in each sample using the relation outlined in \citet{Magdis_c}. Total SFRs are calculated from the rest-frame UV flux (unobscured) and integrated FIR luminosity (obscured). The high-pSFR (diamond with vertical arrow) is an upper limit as the temperature is not well constrained - lower limit is displayed as the horizontal line below the arrow and is derived from the unobscured SFR. The red arrows, without a diamond, display the range for the typical $z\sim3$ LBGs (lower limit from the unobsured SF observed in the UV and upper limit from the observed unobserved + obscured limit). Both assume a stellar mass of 10$^{10.24}$M$_{\odot}$. The green square displays the values for local LBG analogue Haro 11 \citep{Grimes07}. We over plot the observed correlation at $z=0$ \citep{Noeske07}, $z=1$ \citep{Elbaz07}, $z=2$ \citep{Daddi08} and $z=3$ \citep[blue lines,][]{Magdis_c}. Upper and lower boundaries of the $z=3$ distribution are displayed as dashed blue lines. Our subsamples and our limit to the typical $z\sim3$ LBG are consistent with the $z=3$ distribution of IRAC detected LBGs. Errors in the obscured SFR are estimated from our dust SED fitting and are found to be smaller than the plotted symbols. }

\label{fig:SFR_vs_mass}
\end{center} 
\end{figure}

Using our IRAC-inferred stellar masses and total SFRs, we can consider our samples in comparison to the `main sequence'  of  star-forming galaxies. A number of recent studies \cite[$e.g.$][]{Noeske07, Elbaz07,Daddi08, Rodighiero10,Magdis_c} have displayed a  correlation between UV-corrected (total) SFRs and stellar mass in star-forming galaxies out to $z\sim3$. This `main sequence' of star-forming galaxies, predicts increasing star-formation activity with stellar mass (as found in our results). \cite{Magdis_c} determine  this relation for a sample of IRAC detected LBGs $z\sim3$ (see their Figure 7). Figure \ref{fig:SFR_vs_mass} displays the SFR as a function of stellar mass for star-forming galaxies at a range of redshift. We find that our samples are consistent with the \cite{Magdis_c} relation, falling close to their best fit correlation and displaying a similar slope of increasing star-formation activity with stellar mass. We estimate potential errors on the obscured SFR induced by our SED fitting by calculating L$_{FIR}$ (and therefore SFRs) for the range of dust SED temperatures consistent with our data. We find at most 50$\%$ error in our obscured SFRs. These samples are also consistent with the correlation obtained for $sBzK$ at $z\sim2$ \cite{Daddi08}, unsurprising as they have similar characteristics to those of the more massive $z\sim 3$ LBGs. We also plot a range of potential SFRs for our full $z\sim3$ sample. The upper limit of this is constrained by the observed UV SFR + the limit to the obscured SFR derived from our stacked image. The lower limit is constrained by the observed UV SFR alone. We find that our full $z\sim3$ sample falls below the best fit to the $z=3$ main sequence, but is still consistent with the spread of source properties in \cite{Magdis_c}. Clearly with only four data points we can not constrain the SFR-stellar mass relation further. However, it is interesting to note that using a more direct measurement for the total star-formation (UV+FIR) in $z\sim3$ systems, as discussed here, is consistent with the previously obtained correlation. We once again plot the measurement for local LBG analogue Haro 11 \citep[green square, SFR\,$\sim$\,25\,M$_{\odot}$yr$^{-1}$,][]{Grimes07} and find a lower SFR/M$_{*}$ ratio than our detected $z\sim3$ subsamples. However, Haro 11 is still consistent with our full $z\sim3$ sample and the main sequence of star-forming galaxies at $z\sim3$ \citep{Magdis_c}.

\subsection{Testing previous UV-continuum slope predictions}
\label{sec:uv_slope}

In a number of previous studies it has been suggested that deviations from the intrinsic UV continuum spectral slope ($\beta_{int}$) can be used to predict the FIR emission from high redshift sources  \citep[$e.g.$][]{Meurer97,Chapman00,Finkelstein09, Chapman09}. The presence of dust will preferentially attenuate shorter wavelength photons causing the UV spectral slope to be reddened. This attenuated flux will be re-emitted at FIR wavelengths. Hence, there ought to be a correlation between the quantity of flux absorbed in the UV, and that which is emitted in the FIR. However, such an analysis relies on the premise that all deviation from a UV-spectral which is relatively flat in $f_{\nu}$ ($\beta_{int}\sim$-2.0 for a Salpeter IMF) is caused by dust extinction rather than being intrinsic to the source's stellar populations and requires that we know the form of the dust extinction.  Here we can directly test the reliability of these predictions by comparing the rest-frame UV and FIR characteristics of our sub-mm detected samples. Similar analysis has previously been undertaken with FIR observations of lensed $z\sim3$ LBGs and exceptionally FIR bright Lyman-break selected sources \citep[see summary in][]{Chapman09}. These studies find mixed results, with lensed LBGs displaying FIR fluxes which are roughly consistent with the UV spectral slope predictions (within a factor of $\sim2$) and FIR bright LBGs being under predicted by up to a factor of $\sim6$ \citep[WestÐMMD11,][]{Chapman09}. In addition to these, two LBGs (the Cosmic Eye and Cosmic Horseshoe) have had their FIR emission inferred from CO observations \citep{Greve05} and in both cases the CO inferred FIR emission is below the UV-predicted value. 

\begin{figure}
\begin{center}

\includegraphics[scale=0.42]{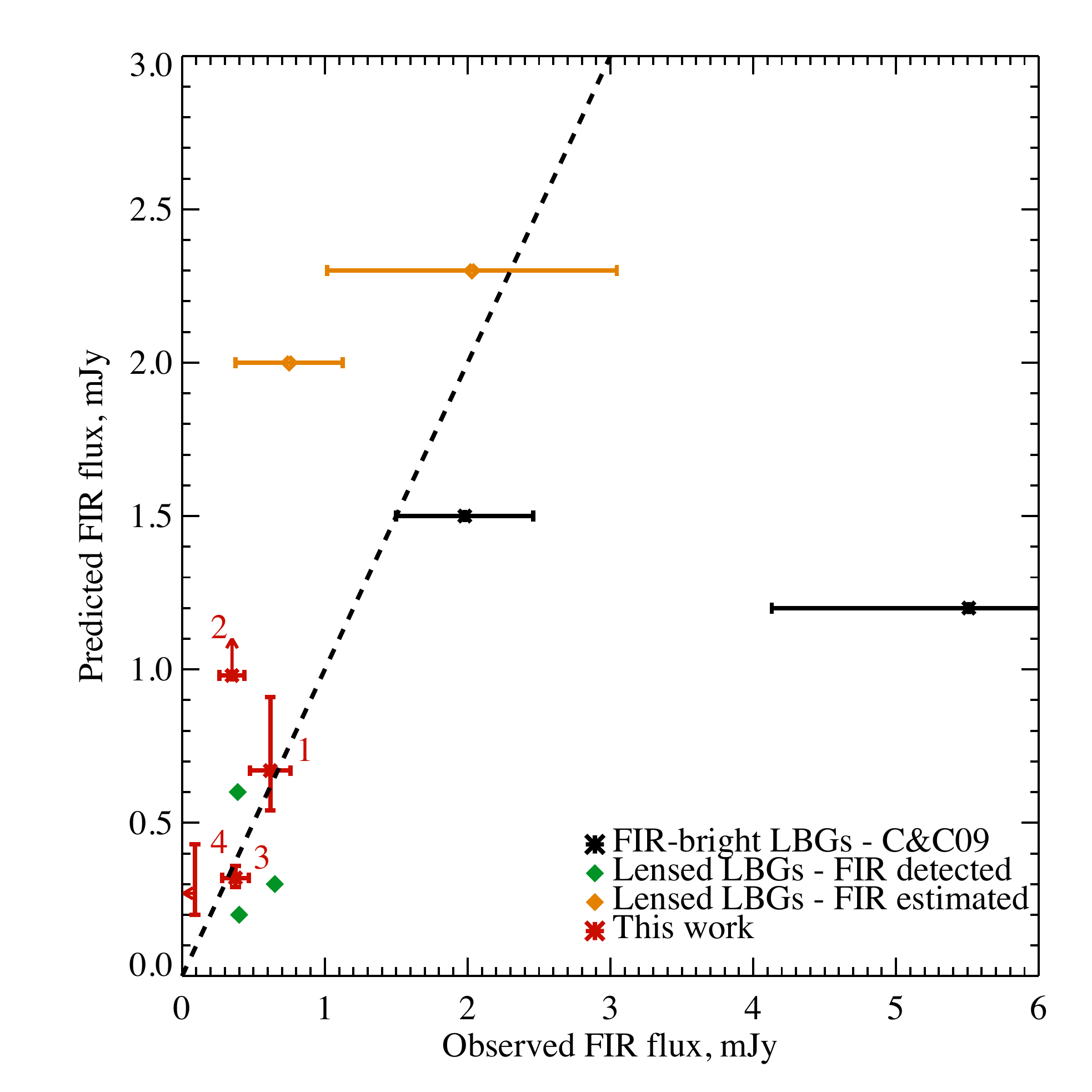}

\caption{Comparison of UV-predicted FIR fluxes with observed FIR fluxes. $z\sim3$ subsamples discussed in this work are displayed as red points labeled: 1. IRAC-22.5, 2. high-pSFR, 3. UV-bright, 4. full $z\sim3$. Green diamonds display lensed LBGs with detectable FIR emission, black stars show exceptionally FIR bright LBGs and orange diamonds display lensed LBGs with FIR fluxes estimated from CO measurements. All points are taken from the summary table in \citet{Chapman09}. The dashed line displays the 1:1 correlation. Our IRAC-22.5, UV-bright and complete $z\sim3$ samples are consistent with the lensed LBGs and have observed FIR fluxes which are well predicted by the UV-continuum slope. However, our high-pSFR sample is not well predicted by the UV slope. }

\label{fig:C&C09}
\end{center} 
\end{figure}

For our $z\sim3$ samples we calculate UV predicted star-formation rates in a similar manner to that discussed in Section \ref{sec:high_SFR} for the high-pSFR sample. We calculate the UV spectral slope ($\beta_{UV}$) in the rest-frame $\sim1250-2250$\AA\ and estimate the total extinction corrected SFR for each of our $z\sim3$ galaxies as in Section \ref{sec:high_SFR}. We then subtract the observed unobscured UV SFR from the total SFR to obtain a predicted FIR SFR (assuming all flux absorbed in the UV is re-emitted at FIR wavelengths). We use this SFR to calculate a predicted FIR luminosity and, for the best-fit temperature in our SED fitting, calculate a predicted 870$\mu$m flux (see Tables \ref{tab:UV_props} and \ref{tab:FIR_props}). Figure \ref{fig:C&C09} shows the observed against predicted FIR fluxes for our $z\sim3$ subsamples (red points). We over-plot the values obtained for lensed LBG (green diamonds), FIR bright LBGs (black stars) and LBGs with FIR fluxes estimated from CO measurements (orange diamonds) - all taken from the summary table in \cite{Chapman09}. The dashed line displays a 1:1 correlation. We find the observed fluxes from our IRAC-22.5 and UV-bright samples are consistent with the UV-continuum slope predictions and the limit from our full $z\sim3$ sample is consistent with being within a factor of two UV predicted values - similar to the lensed sources, suggesting that the these highly magnified, serendipitous source may be representative of the general population (note that the range of UV predicted fluxes for the full sample represents a temperature range from 30-40K). However, in our high-pSFR sample the UV prediction overestimates the observed FIR flux by at least a factor of $\sim$3 indicating that at least one of the assumptions used in predicting the total SFR (the intrinsic UV slope, form of the dust correction) is likely to be incorrect for this sample.

Interestingly we find that our high-pSFR and UV-bright samples have distinctly different predicted FIR fluxes, while their observed fluxes are comparable. These samples are almost identical in mean rest-frame UV (1600\AA) magnitude, suggesting that they share similar unobscured SFRs, and have comparable stellar masses, but display significantly different UV spectral slopes ($\beta_{UV}$=-1.44 and $\beta_{UV}$=-1.76 respectively, as noted previously). Following the previously suggested scenario, this shallower spectral slope should be indicative of greater FIR emission (for the same UV magnitude, as is the case here). While the dust temperature of the high-pSFR sample is not well constrained, the non detections in the $Herschel$ bands predict an upper limit to the temperature of $\sim34$K. Extrapolating further, to obtain a SFR which is consistent with that predicted from the UV continuum slope, we would require a dust temperature of $\sim42$K. This is clearly ruled out by the $Herschel$ non-detections. With the caveat that these values are constrained from the mean of a large sample (and hence will not show source to source variation), this result suggests that potentially the FIR emission from these sources is not directly correlated to the UV-spectral slope, and may be dependant on other factors - thereby implying that an assumption that all sources have an intrinsic spectral slope of $\beta_{UV}$=-2.0 may be invalid, or that the assumptions in correcting for dust extinction (primarily the extinction law) are incorrect. Following this our high-pSFR sample is not in fact a sample of the most highly star-forming LBGs when we consider the FIR emission directly.    

The failure of the dust correction to predict the FIR luminosity of the high-pSFR sample does not necessarily indicate that the correction procedure is inappropriate for most objects (it appears appropriate for the IRAS-22.5 and UV-bright samples). The method of selection for the high-pSFR sample means it contains objects with the largest UV-slope correction, while at the same time having relatively high observed UV fluxes - resulting in the largest correction for obscured star formation. It is possible  this combination of properties may indicate that the sources have red UV slopes for reason other than dust attenuation (possibly a wide range of ages, star-formation histories or both). Clearly, there are $z\sim3$ objects with higher SFRs than the rest of our samples (such as typical SMGs), but these objects may be so reddened that they miss our $V-$band or $V-R$ cuts  and therefore are not selected by us. 

\subsection{Correlations with stellar mass - low mass SMG type sources?}
\label{sec:mass}

\begin{figure}
\begin{center}

\includegraphics[scale=0.42]{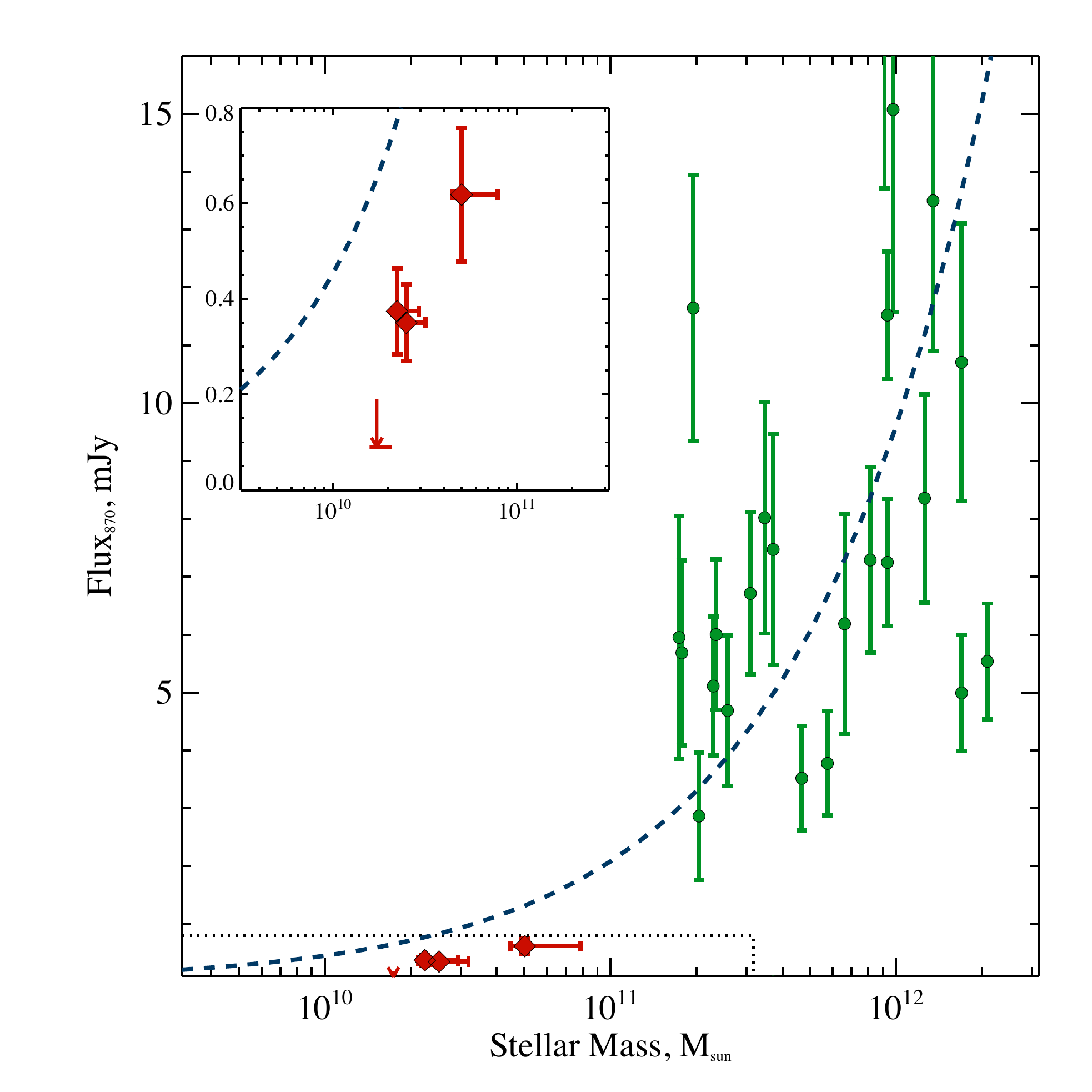}

\caption{The FIR flux against stellar mass for our $z\sim3$ samples (red diamonds). Over plotted are  $z>2.5$ submm galaxies taken from \citet{Michalowski10} - fluxes scaled so all sources are at z$\sim$3 (green circles). The orange line displays the best fit to the submm galaxy points, while the blue dashed line displays the predicted correlation from hydro dynamical simulations of isolated disk galaxies at $z\sim3$ \citep{Hayward13}. The inset displays a zoomed in region bounded by the dotted line.}

\label{fig:flux_v_mass}
\end{center} 
\end{figure}

Our results imply a correlation between stellar mass and FIR emission. While this is similar to the `main sequence' of star-forming galaxies, it does not contain any correlation of stellar mass with UV emission and simply relates the stellar mass to the observed FIR emission directly.  Figure \ref{fig:flux_v_mass} (inset) displays observed FIR flux against stellar mass for our $z\sim3$ samples. We find a deceasing 870$\mu$m flux with decreasing stellar mass between our four samples, indicating that irrespective of UV spectral slope and UV star-formation rates, more massive systems have larger FIR fluxes. Additionally, in the full Figure \ref{fig:flux_v_mass}, we compare our samples with the distribution of $z>2.5$ submm galaxies taken from \citep{Michalowski10}, with fluxes scaled to a flux assuming the source is at $z=3$.  We also over plot the predicted correlation derived from hydrodynamical simulations of isolated disk galaxies at $z\sim3$ \cite[which may be appropriate for both SMGs and LBGs - blue dashed line,][]{Hayward13}.  We find that this model over-predicts the 870$\mu$m flux from all of our samples, potentially due to the fact that the Hayward et al models assume that the galaxy is completely obscured (this will not be the case for our rest-frame optically selected galaxies). 

\begin{figure}
\begin{center}

\includegraphics[scale=0.40]{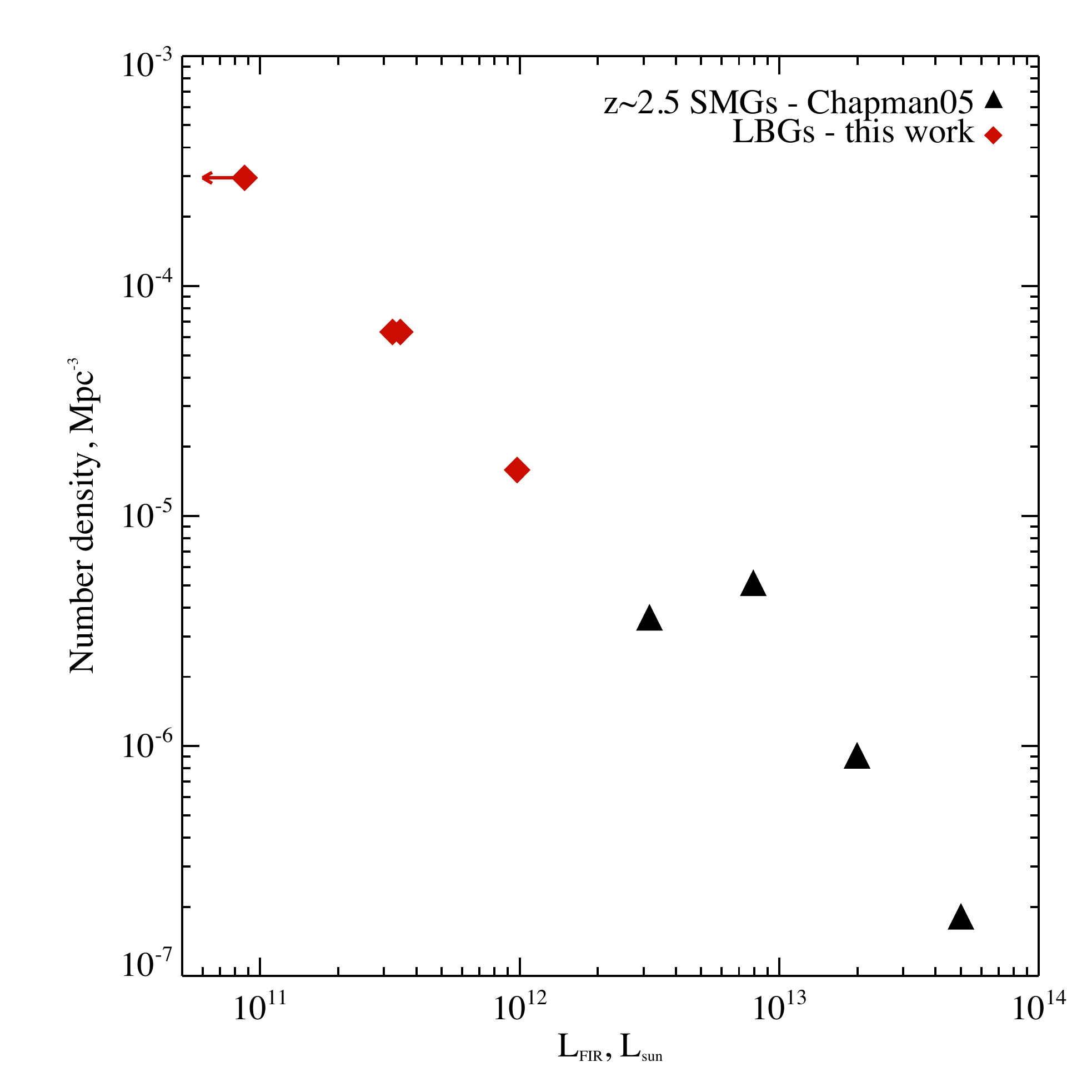}

\caption{The FIR luminosity function for $z\gtrsim2.5$ sources. Black triangles show the luminosity function for $z\sim2.5$ SMGs taken from \citet{Chapman05}. The red diamonds display the best fit FIR luminosity against number density for our $z\sim3$ samples. Our results are consistent with an extension of the FIR luminosity function for $z\sim2.5$ SMGs to fainter FIR luminosities. }

\label{fig:lum_func}
\end{center} 
\end{figure}

A simple interpretation of this is that the most massive LBGs at $z\sim3$ are simply the low mass end of the submm mass function at $z\sim3$. This is consistent with the assertion of \cite{Michalowski12}, that SMGs are not pathological objects but the top end of the `typical' galaxy mass function at high redshifts. We have essentially highlighted a sample of more `typical' $z\sim3$ galaxies (than SMGs) and shown their FIR characteristics to be an extension of those found in submm luminous sources, albeit at lower masses. Our populations display number number densities of $\sim$few$\times10^{-3}$\,Mpc$^{-3}$ for our full $z\sim3$ sample, $\sim$few$\times10^{-4}$\,Mpc$^{-3}$ for the UV-bright sample and $\sim10^{-5}$\,Mpc$^{-3}$ for the stellar mass selected IRAC-22.5 sample. These are significantly higher that those derived for $z\sim2.5$ spectroscopically confirmed SMGs \citep{Chapman05}, but are fully consistent with an extension of the lower luminosity end of the $z\sim2.5$ FIR luminosity function (Figure \ref{fig:lum_func}).      

This adds weight to the argument of a continuous `main sequence' of star-forming galaxies at high redshift, with simply the fraction of obscured star-formation increasing with stellar mass - starting with mostly unobscured `typical' LBGs, through the massive IRAC-22.5 LBGs of our sample, to SMGs. Clearly, identifying systems in the main sequence `void' between LBGs and SMGs (around M$_{*}\sim10^{11}$M$_{\odot}$) will be problematic. As systems become more extincted, they will be less UV luminous and drop out of Lyman break selected samples. However, these galaxies may not be FIR luminous  enough to be identified in deep submm surveys. Figure \ref{fig:flux_v_mass} suggests that we may hope to identify a sample of yet more massive ($>10^{11}$M$_{\odot}$) $z\sim3$ galaxies with FIR fluxes in the 1.0-3.0\,mJy range. Such systems would potentially populate the `main sequence' between SMGs and LBGs, displaying both significant unobscured and obscured star-formation, with the faction of obscured material increasing with stellar mass. Key targets for the identification of such sources will be the most massive systems at $z\sim3$, which nonetheless fail to reach the detection limits in current deep submm surveys.

\section{Summary and Conclusions} 

We have selected robust samples of LBGs over a range of epochs in the ECDF-S selecting 922, 68 and 20 LBGs at $z\sim3, 4,$ and 5 respectively. We produce composite images of these full samples at $870\mu$m and do not obtain detections. At $z\sim3$, where our sample size allows, we selected subsamples of LBGs on stellar mass, predicted (extinction corrected) total SFRs and UV-luminosity. We produce composite images of these subsamples at multiple wavelengths spanning the dust emission peak at $z\sim3$ and constrain the dust SED of our stellar mass and UV-selected samples. Using this we calculate best-fit dust temperatures (33-41\,K), FIR luminosities ($10^{11.5-12.0}$L$_{\odot}$) and obscured SFRs ($\sim$40-250\,M$_{\odot}$\,yr$^{-1}$ including errors), and find that a significant fraction of their star-formation is obscured - up to 80$\%$. We compere the FIR properties of our samples with other galaxy populations and find that our stellar mass selected sample shares some characteristics with SMGs at the same epoch, albeit at lower masses and FIR luminosities - they have similar dust fractions and temperatures ($\sim$40\,K). We calculate total (UV+FIR) SFRs for our samples and find that they all fall on the main sequence of star-forming galaxies at high redshift, suggesting that direct measurements of the obscured SFRs in these systems leads to total SFRs which are consistent with previous predictions (which do not use detections in the FIR). 

A number of these previous predictions use the UV-continuum slope, $\beta_{UV}$, to estimate FIR fluxes from high redshift sources - essentially assuming any deviation from in intrinsic spectral slope of -2 is due to dust attenuation. Here we test those predictions directly by comparing the UV-slope inferred 870$\mu$m emission with that observed directly in our composite images. We find that for our full $z\sim3$ sample, stellar mass selected sample and UV-bright sample, the predictions match the true FIR emission relatively well (within a factor of two). However, for those LBGs predicted to have the highest total SFR, generally those bright in the UV but with the largest deviation from a flat ($\beta=-2$ )UV spectral slope, this correction does not appear to work. The true 870$\mu$m flux of these sources is similar to samples with the same observed UV luminosity, but with flatter UV spectral slopes and a factor of $\sim 3$ times lower than predicted. This does not  indicate that this method is inappropriate for the majority of sources, just those with the most extreme corrections. It appears that the most highly star forming systems (with rates of  several $100$M$_{\odot}$yr$^{-1}$) do not appear in our sample, presumably as they fail to make our LBG magnitude and/or colour cuts due to their reddening.   

We predict rest-frame optically derived stellar masses for our samples and display a potential correlation between stellar mass and 870$\mu$m flux (consistent with the non-detections in our full samples at all redshifts). The most simple interpretation of this is that our submm detected samples are simply the low mass end of the submm mass function at $z\sim3$. We also compare the number density of our samples to those for SMGs at $z\sim2.5$ and show that our sample may well represent the lower luminosity end of the high redshift FIR luminosity function.  Assuming this stellar mass to FIR flux correlation extends to lower masses, and the same dust SED is applicable in higher redshift systems, observations much deeper than those obtainable by single dish observatories will be required to individually detect the typical LBG at $z\gtrsim3$.

\section*{Acknowledgements}
This paper makes use of data obtained for APEX program IDs 078.F-9028(A), 079.F-9500(A), 080.A-3023(A), and 081.F-9500(A).  APEX is operated by the Max-Planck-Institut fur Radioastronomie, the European Southern Observatory, and the Onsala Space Observatory. In addition, we made use of data from HerMES project (http://hermes.sussex.ac.uk/). HerMES is a Herschel Key Programme utilizing Guaranteed Time from the SPIRE instrument team, ESAC scientists and a mission scientist. HerMES is described in Oliver et al. 2012. We would also like to thank Chris Hayward for his helpful comments.

\end{document}